\documentclass[review,authoryear,12pt]{elsarticle}
\usepackage{graphicx}
\usepackage[tight,footnotesize, nooneline]{subfigure}
\usepackage[hidelinks]{hyperref} % pdf hyperlinks

\usepackage{amssymb}
\usepackage{amsmath}

\usepackage{rotating}

\usepackage{epstopdf}

\usepackage{lineno}
 
\usepackage{multirow}

\journal{arXiv}

\newcommand\copyrighttext{%
   \textcopyright\ 2016. This manuscript version is made available under the CC-BY 3.0 license \url{https://creativecommons.org/licenses/by/3.0/}}

\begin{document}

\begin{frontmatter}

%% Title


\title{The effect of temperature dependent tissue parameters on acoustic radiation force induced displacements}

%% Authors and addresses/affiliations

%% use the corref command within \author for corresponding author footnotes; note the corresponding author can be any of the authors, but one author must be designated; here the third author has been arbitrarily designated as the corresponding author as an example.



\author[Affil1]{Visa Suomi \corref{cor1}}
\author[Affil2]{Yang Han}
%\author[Affil1]{David Edwards}
\author[Affil2]{Elisa Konofagou}
\author[Affil1]{Robin Cleveland}
%\author[Affil2]{Second Author}
\address[Affil1]{Department of Engineering Science, University of Oxford, Parks Road, Oxford, OX1 3PJ, UK}
\address[Affil2]{Department of Biomedical Engineering, Columbia University, 1210 Amsterdam Avenue, 351 Engineering Terrace, New York, NY 10027, USA}
%\address[Affil2]{Affiliation address 2}
% Replace capitalized text with the appropriate information (use standard capitalization rules for your text, not all capitals.
\cortext[cor1]{Corresponding Author: Visa Suomi, Institute of Biomedical Engineering, Old Road Campus Research Building, University of Oxford, Oxford OX3 7DQ, UK; Email, visa.suomi@eng.ox.ac.uk; Phone, +44 (0) 1865 617660}

\begin{abstract}
Multiple ultrasound elastography techniques rely on acoustic radiation force (ARF) in monitoring high-intensity focused ultrasound (HIFU) therapy. However, ARF is dependent on tissue attenuation and sound speed, both of which are also known to change with temperature making the therapy monitoring more challenging. Furthermore, the viscoelastic properties of tissue are also temperature dependent, which affects the displacements induced by ARF. The aim of this study is to quantify the temperature dependent changes in the acoustic and viscoelastic properties of liver and investigate their effect on ARF induced displacements by using both experimental methods and simulations. Furthermore, the temperature dependent viscoelastic properties of liver are experimentally measured over a frequency range of 0.1-200 Hz at temperatures reaching 80 $^{\circ}$C, and both conventional and fractional Zener models are used to fit the data.

The fractional Zener model was found to fit better with the experimental viscoelasticity data with respect to the conventional model with up to two magnitudes lower sum of squared errors (SSE). The characteristics of experimental displacement data were also seen in the simulations due to the changes in attenuation coefficient and lesion development. At low temperatures before thermal ablation, attenuation was found to affect the displacement amplitude. At higher temperature, the decrease in displacement amplitude occurs approximately at 60-70 $^{\circ}$C due to the combined effect of viscoelasticity changes and lesion growth overpowering the effect of attenuation. The results suggest that it is necessary to monitor displacement continuously during HIFU therapy in order to ascertain when ablation occurs.
\end{abstract}


\end{frontmatter}

\copyrighttext

%% Do not remove the page break here.
\pagebreak

%\linenumbers

%% MAIN TEXT INSTRUCTIONS

%% For all sections, subsections, and subsubsections, use the '*' to remove numbering, as demonstrated below.

%% Commands for figures and tables should not be included in the main body of the submitted version of this file (e.g. the figure and tabular environments).  Figure captions should be listed in this file, as shown below.  Tables and Table captions should be listed as a separate section at the end of this file, as shown below.  Many authors may wish to include figures and tables within the main text of their document will preparing their manuscript.  This may be done, however please comment out any of the lines prior to submission.

%% Because the Elsevier editorial process does not allow for the figure and tabular environments in the submitted document, you will be unable to use autonumbering (i.e. \label and \ref) for figures and tables. 

%%  If long equations are used in the document, authors should use a two column format to make sure that the equations will break at approximately the right places.  To do this, replace the class option 'review' with the following two class options '3p' and 'twocolumn'.  Keep in mind that the column width produced in '3p' is slightly narrower than the final printer version.  After inserting the appropriate line breaks in your equation, change the '3p' option back to 'review'.

%% For citations, use the commands \citep and \citet

%% BEGIN MAIN TEXT

%%%%%%%%%%% INTRODUCTION
\section*{Introduction}

One of the main barriers for the wide spread use of high-intensity focused ultrasound (HIFU) has been the lack of robust, reliable and cost-efficient monitoring methods. Monitoring HIFU therapy requires real-time imaging which can provide feedback as to the state of the tissue either in the form of temperature data or by determining changes in the physical properties of tissue. Ultrasound can be used to provide temperature information, but this has been found to be accurate only at hyperthermic temperatures below 50 $^{\circ}$C \citep{lewis2015thermometry}. In the case where lesions are echogenic, B-mode imaging can be used to monitor HIFU. However, in the absence of cavitation lesions are not always visible \citep{bush1993acoustic, rabkin2005hyperecho}.

One tissue property that has been shown to change significantly and irreversibly after thermal ablation is tissue stiffness \citep{wu2001assessment, sapin2010temperature}. The increase in stiffness has been shown to be reversible at temperatures below 60 $^{\circ}$C \citep{wu2001assessment}, and irreversible once temperatures exceed 60 $^{\circ}$C due to collagen denaturation \citep{wu2001assessment, lepetit2000modelling}. This suggests that measurements of stiffness can be used to distinguish thermally ablated tissue \citep{varghese2002elastographic, varghese2003elastographic}.

Ultrasound elastography imaging is a non-invasive method for measuring mechanical properties of tissue \citep{palmeri2011acoustic, gennisson2013ultrasound}. It is a widely used medical imaging technique, which has several clinical applications \citep{cosgrove2013efsumb}. Ultrasound elastography can employ quasi-static approaches \citep{ophir1991elastography} or create deformations through acoustic radiation force (ARF), for example, acoustic radiation force impulse (ARFI) imaging \citep{lizzi2003radiation}, supersonic shear imaging (SSI) \citep{bercoff2004monitoring} and harmonic motion imaging (HMI) \citep{maleke2010vivo}.

ARF results when propagating ultrasound waves transfer their momentum to the surrounding medium by the processes of absorption and reflection \citep{torr1984acoustic}. The magnitude of the force per unit volume $F_{V}$ (kg$\cdot$s$^{-2}\cdot$m$^{-2}$) is given by equation \citep{nyborg1965acoustic, torr1984acoustic, starritt1991forces}:
\begin{equation}
\label{eq:ARF}
F_{V} = \frac{2 \alpha_{\mathrm{abs}} I_{\mathrm{TA}}}{c}
\end{equation}
where $\alpha_{\mathrm{abs}}$ (Np$\cdot$m$^{-1}$) is the absorption of the medium, $I_{\mathrm{TA}}$ (W$\cdot$m$^{-2}$) is the temporal average intensity of the ultrasound field at a given location and $c$ (m$\cdot$s$^{-1}$) is the sound speed in the medium. The resulting deformation caused by ARF is described by \citep{calle2005temporal}:
\begin{equation}
\rho \frac{\partial^{2} u_{i}}{\partial t^{2}} =  \frac{\partial T_{ij}}{\partial x_{j}} + F_{Vi}
\end{equation}
where $u_{i}$ (m) is the displacement in the direction of $x_{i}$ (m), $\rho$ (kg$\cdot$m$^{-3}$) is the density, $t$ (s) is time, $T_{ij}$ (Pa) is the stress tensor and $F_{Vi}$ (kg$\cdot$s$^{-2}\cdot$m$^{-2}$) is the radiation force in the direction of $x_{i}$. The temperature dependence of the deformation is affected by the changes in the sound speed, attenuation and elasticity, and therefore, extracting temperature from measured deformation is challenging.

The first aim of this study is to quantify the temperature dependence of acoustic and viscoelastic tissue parameters affecting ARF induced displacements. This is conducted by summarising the existing data on acoustic and elastic properties available in the literature. Furthermore, the temperature dependence of viscoelastic properties in \textit{ex vivo} liver is experimentally measured and both conventional and fractional Zener model parameters are fit to the data. The second aim is to demonstrate how HMI can be used to monitor HIFU thermal ablation in \textit{ex vivo} liver. Finally, the characteristics of the experimental HMI data are modelled by viscoelastic finite element method (FEM) simulations using temperature dependent acoustic and viscoelastic parameters.

\subsection*{Temperature dependence of acoustic properties in liver}

According to (\ref{eq:ARF}) the magnitude of the ARF is directly dependent on absorption of tissue. However, in soft tissue absorption is the main factor contributing to attenuation \citep{lyons1988absorption} and in calculating ARF the attenuation coefficient is employed.

Temperature dependence and the effect of thermal ablation on liver ultrasonic attenuation has been extensively studied by several authors \citep{bamber1979ultrasonic, gammell1979temperature, bush1993acoustic, damianou1997dependence, gertner1997ultrasound, techavipoo2002temperature, techavipoo2004temperature, choi2011changes, ghoshal2011temperature, zderic2004attenuation, kemmerer2012ultrasonic, jackson2014nonlinear}.

\begin{figure*}[!htbp]
\begin{center}
    \subfigure[]
    {
        \includegraphics[width=1\textwidth]{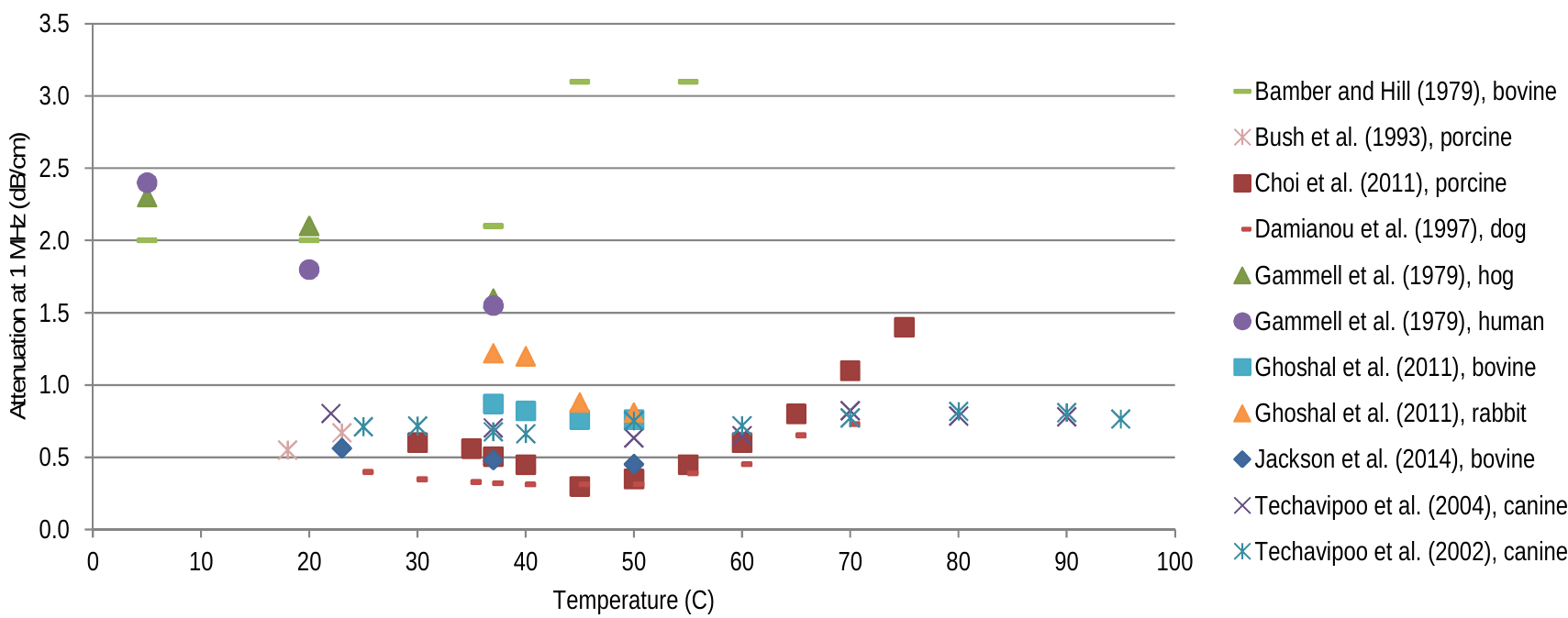}
    }
    \\
    \subfigure[]
    {
        \includegraphics[width=1\textwidth]{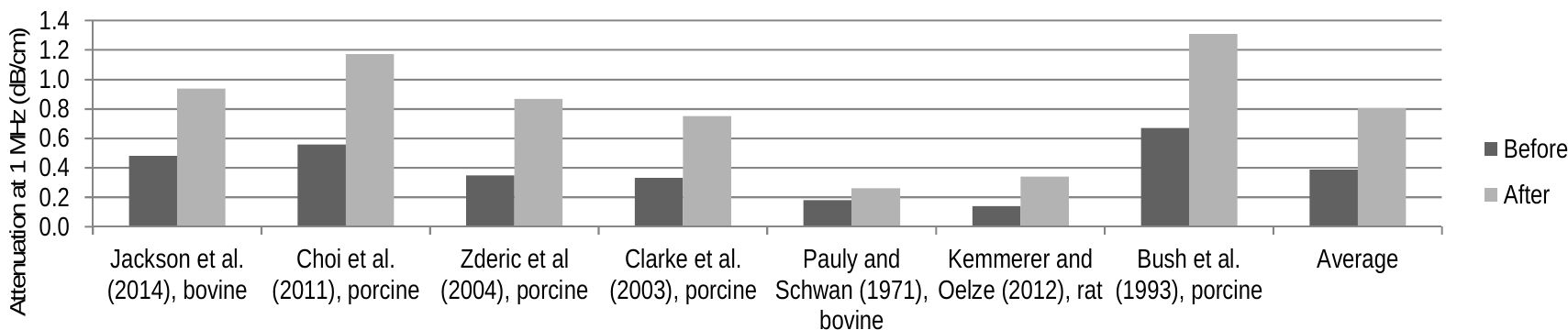}
    }
    \\
    \subfigure[]
    {
        \includegraphics[height=3cm]{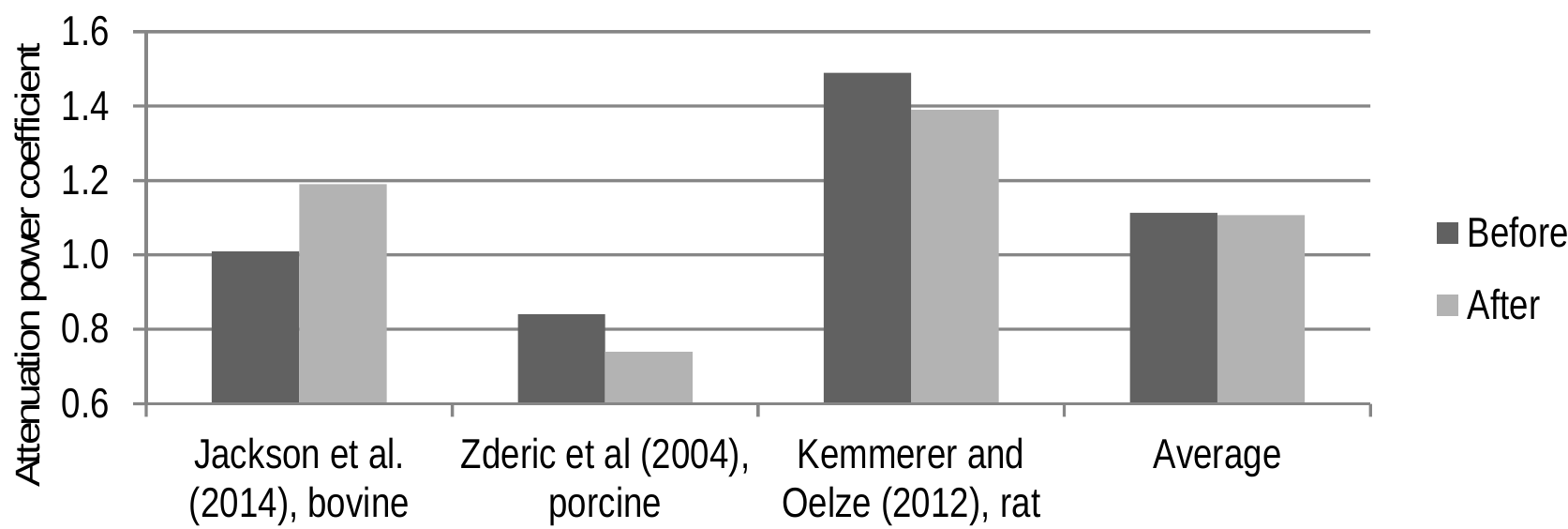}
    }
\end{center}
\caption{Data collected from published literature showing the changes in (a) attenuation coefficient with temperature in liver. (b) Attenuation coefficient and (c) attenuation power law coefficient before and after thermal ablation in liver.}
\label{fig:literature_att}
\end{figure*}

Figure \ref{fig:literature_att}(a) shows the temperature dependence of attenuation coefficient from various studies. The scatter in the data is most likely due to differences in measurement techniques and biological variability. Figures \ref{fig:literature_att}(b) and (c) show the effect of thermal ablation on the attenuation coefficient and the attenuation power law respectively. On average the attenuation coefficient increases by approximately 110\% after thermal ablation which suggests more than a two-fold increase in ARF. The attenuation power law decreases by less than 0.3\% on average.

\begin{figure*}[!htbp]
\begin{center}
    \subfigure[]
    {
        \includegraphics[width=1\textwidth]{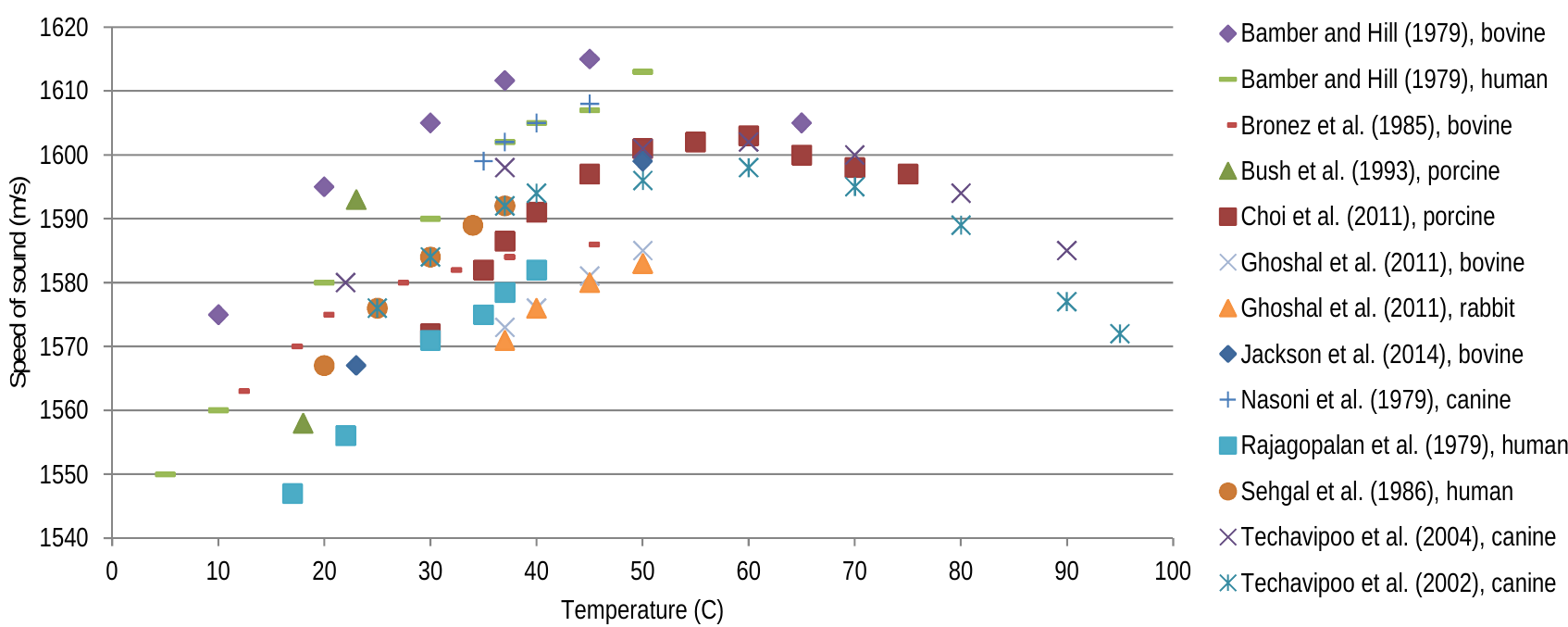}
    }
    \\
    \subfigure[]
    {
        \includegraphics[width=1\textwidth]{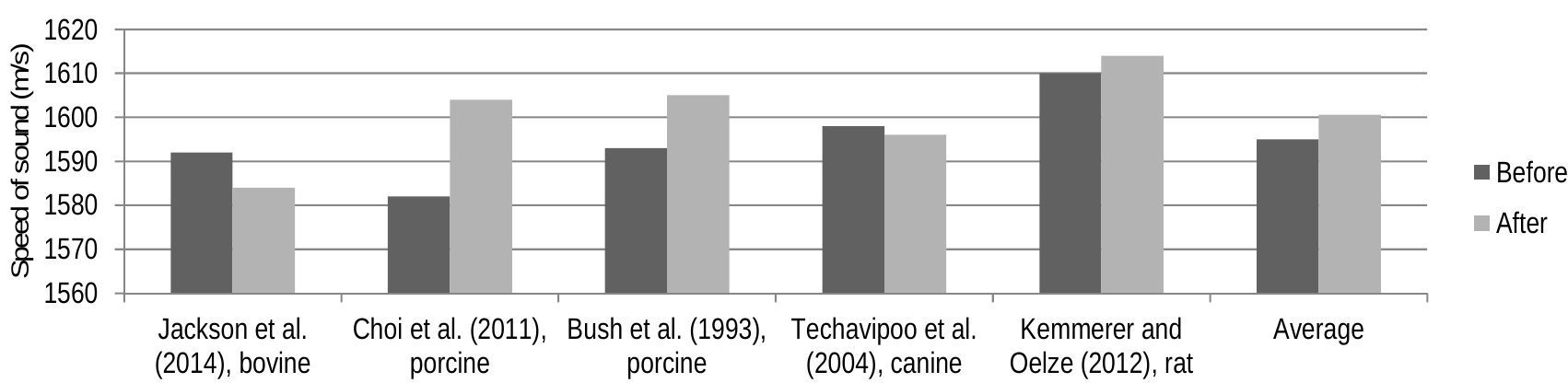}
    }
\end{center}
\caption{Data collected from published literature showing the changes in sound speed in liver (a) with temperature and (b) before and after thermal ablation.}
\label{fig:literature_sos}
\end{figure*}

The magnitude of ARF is also affected by sound speed and its temperature dependence has been investigated \citep{bamber1979ultrasonic, nasoni1979vivo, rajagopalan1979variation, bronez1985measurement, sehgal1986measurement, bush1993acoustic, techavipoo2002temperature, techavipoo2004temperature, choi2011changes, ghoshal2011temperature, jackson2014nonlinear}. The published data on sound speed temperature dependence in liver is summarised in Figure \ref{fig:literature_sos}(a). In general the sound speed increases with a maximum occurring at temperature in the range 50-70 $^{\circ}$C after which a decrease is seen. The reported data on the effect of thermal ablation on sound speed in liver is shown in Figure \ref{fig:literature_sos}(b) \citep{jackson2014nonlinear, choi2011changes, bush1993acoustic, techavipoo2004temperature, kemmerer2012ultrasonic}. On average the effect post-ablation is small with an increase in sound speed of 0.4\%.

\pagebreak

\subsection*{Temperature dependence of elastic properties in liver}

\citet{sapin2010temperature} studied the temperature dependence of shear modulus in \textit{ex vivo} liver. They reported the shear modulus to be approximately constant up to 45 $^{\circ}$C after which an exponential increase was measured with a maximum modulus of 50 kPa at the temperature of 67 $^{\circ}$C (1250\% increase). The stiffness changes at temperatures above 45 $^{\circ}$C were found to be irreversible. This is due to protein denaturation and tissue necrosis forming a coagulated region which consequently leads to increase in stiffness \citep{lepetit2000modelling, wu2001assessment, sapin2010temperature}. These changes are present also at temperatures used in HIFU therapy and hence cause significant changes in the deformation measured by ultrasound elastography modalities. 

\begin{figure*}[!htbp]
\begin{center}
    {
        \includegraphics[height=4cm]{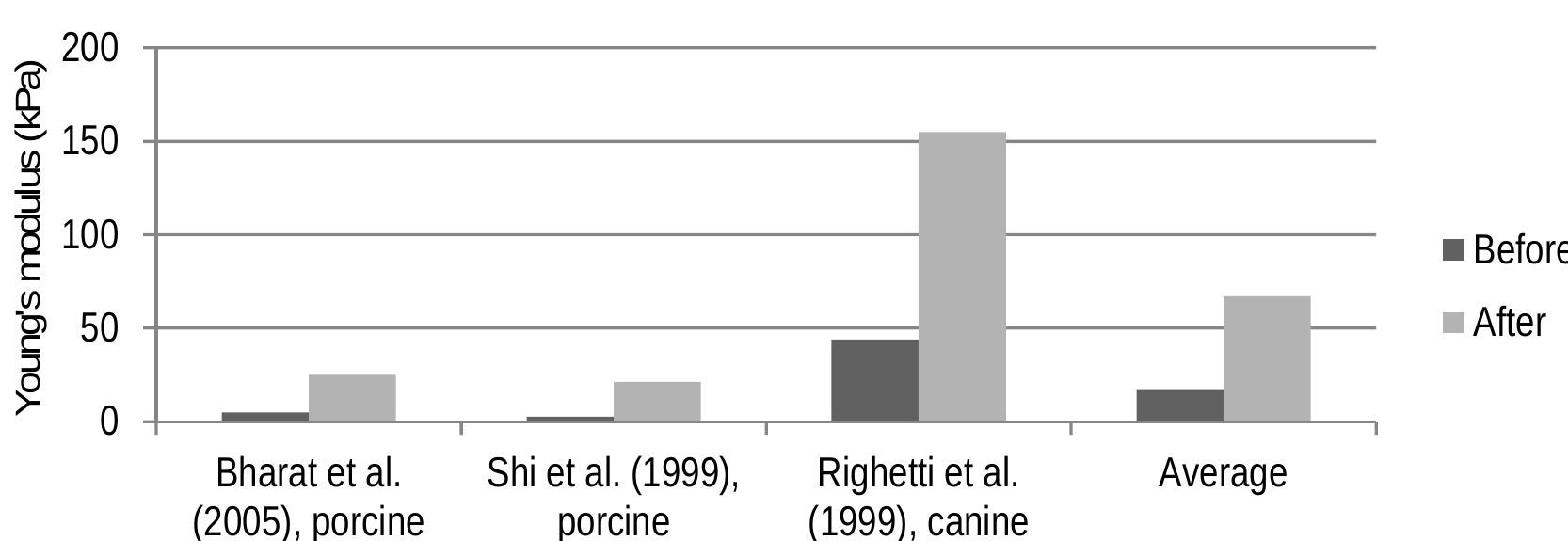}
    }
\end{center}
\caption{Data collected from published literature showing the changes in Young's modulus values before and after thermal ablation in liver.}
\label{fig:literature_ym}
\end{figure*}

Figure \ref{fig:literature_ym} shows the experimental data of Young's modulus before and after thermal ablation \citep{bharat2005monitoring, shi1999detection, righetti1999elastographic} with irreversible changes varying from a four-fold to eight-fold increase.

\subsection*{Temperature dependence of viscoelastic properties in liver}

Typically tissue elastic modulus measurements are conducted using either stress relaxation or creep tests which quantify elastic parameters at slow rates of loading (i.e., several seconds). However, in ultrasound elastography modalities, such as ARFI and HMI, the tissue excitation periods are on the order of milliseconds or less in which case it is necessary to have experimental data over a wider time scale. This can be accomplished by dynamic mechanical analysis (DMA) which is able to characterise the viscoelastic behaviour of tissue in the frequency domain. In DMA the storage (elastic) and loss (viscous) modulus are measured over a range of frequencies.

\citet{kiss2004viscoelastic, kiss2009investigation} used DMA to characterise the viscoelastic properties of thermally ablated lesions in \textit{ex vivo} liver. It was found that the magnitude of the dynamic modulus increased with ablation temperature reaching a local maximum at around 70-75 $^{\circ}$C. In another experiment the magnitude of dynamic modulus continuously increased up to the maximum measurement temperature at 90 $^{\circ}$C. 

The results suggest that the changes in the viscoelastic properties of liver will have a significant effect on the deformations in ultrasound elastography modalities which rely on short duration or frequency specific ultrasound exposures. The temperature dependence of dynamic modulus is yet to be studied which will be addressed in this article together with the fitting of conventional and fractional Zener models. Simpler Maxwell and Kelvin-Voigt models are often used in viscoelasticity studies, but they are lacking the ability to predict both creep-recovery and stress relaxation, and therefore, they are only valid for a very limited set of materials. Zener model (aka standard linear solid model) is a three-parameter model capable of describing both of these general features of viscoelastic behaviour. Due to the significant changes in tissue viscoelasticity in thermal ablation, Zener model is considered more accurate in predicting this behaviour.

\pagebreak

%%%%%%%%%%% MATERIALS AND METHODS
\section*{Materials and methods}

\subsection*{Curve fitting to temperature dependent attenuation coefficient data}

\begin{figure}[!htbp]
\begin{center}
\includegraphics[width=1\textwidth]{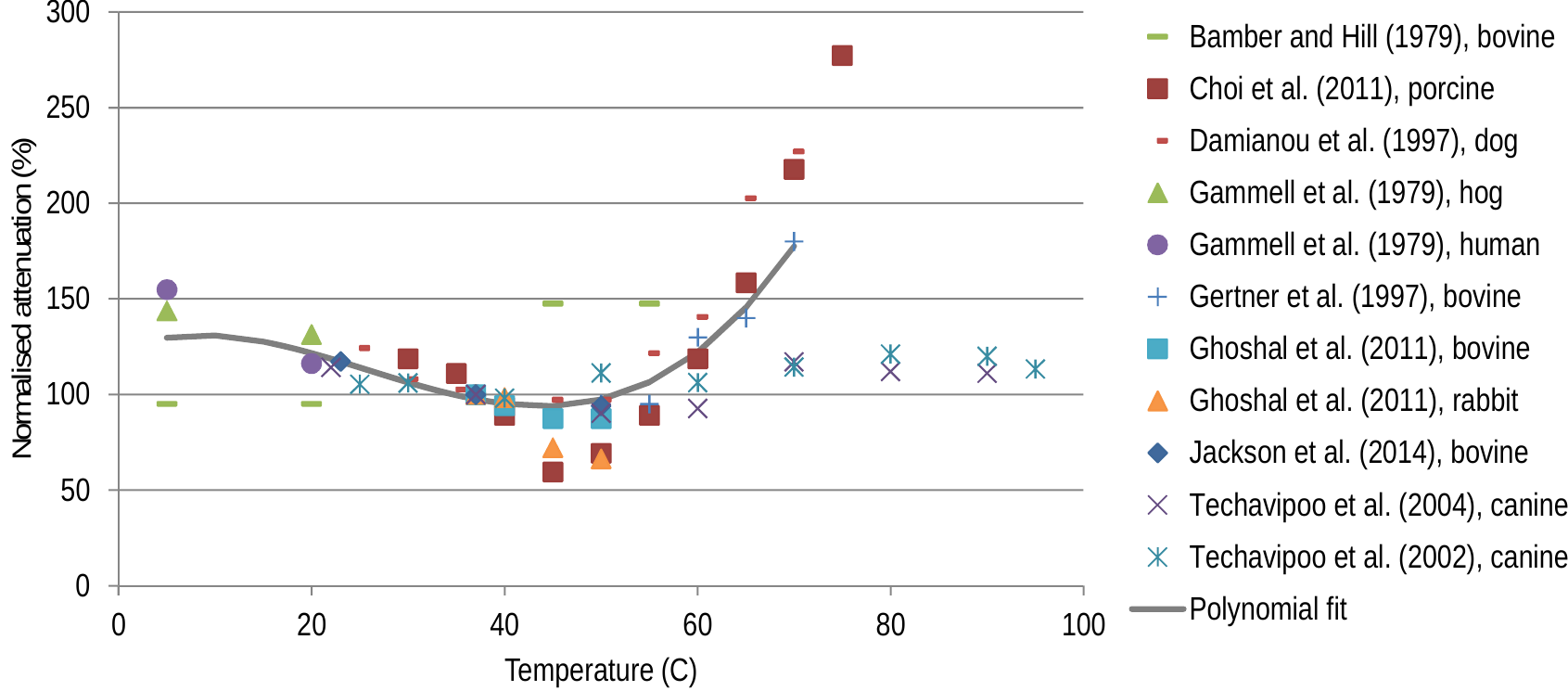}
\end{center}
\caption{Data collected from the literature showing the changes in normalised (to 37 $^{\circ}$C) attenuation coefficient in liver with temperature. Fourth order polynomial function was fit to the normalised data with parameters: $p_{0}$ = 1.23E+02, $p_{1}$ = 1.99E+00, $p_{2}$ = $-$1.41E$-$01, $p_{3}$ = 1.94E$-$03, $p_{4}$ = $-$2.55E$-$06.}
\label{fig:literature_att_norm}
\end{figure}

The acoustic property that showed the strongest temperature effect was the attenuation coefficient, with over a two-fold increase, and therefore a model for the temperature dependence was developed by curve fitting to the published data. The attenuation coefficient data from different authors was scattered. However, after normalising to the value at 37 $^{\circ}$C the data was more consistent (see Figure \ref{fig:literature_att_norm}). After normalisation a fourth order polynomial was fit to the data using a nonlinear least squares method. The fourth order polynomial function was chosen in order to catch the decreasing attenuation around 45 $^{\circ}$C before the subsequent increase.

The function was fit over the temperature range 5-70 $^{\circ}$C because of the large deviation in the attenuation coefficient values above 70 $^{\circ}$C. The fourth order polynomial function is defined by the equation:
\begin{equation}
f_{\textrm{pol}}(T) = p_{0} + p_{1}T + p_{2}T^2 + p_{3}T^3 + p_{4}T^4
\end{equation}
where $T$ ($^{\circ}$C) is the temperature and $p_{0}$-$p_{4}$ are the fitting parameters. The polynomial fit is presented in Figure \ref{fig:literature_att_norm} with the corresponding fitting parameters.

%\begin{table}[!htbp]
%  \centering
%  \caption{Polynomial fitting parameters for the temperature dependence of normalised attenuation in liver.}
%    \begin{tabular}{llllll}
%    \hline
%    \multicolumn{1}{c}{$p_{4}$} & \multicolumn{1}{c}{$p_{3}$} & \multicolumn{1}{c}{$p_{2}$} & \multicolumn{1}{c}{$p_{1}$} & \multicolumn{1}{c}{$p_{0}$} \\
%    \hline
%    $-$2.55E$-$06 & 1.94E$-$03 & $-$1.41E$-$01 & 1.99E+00 & 1.23E+02 \\
%    \hline
%    \end{tabular}
%  \label{tab:att_polynomial}
%\end{table}

\subsection*{Measurement of dynamic modulus}

The temperature dependent dynamic modulus of liver was measured using DMA (Q800, TA Instruments, New Castle, DE, USA) in a frequency range from 0.1 to 200 Hz. A total of 5 samples were cut from freshly obtained (within 48 hours of slaughter) bovine liver. Each sample was cut to a rectangle shape with a size of approximately 20 mm $\times$ 20 mm $\times$ 10 mm and placed in a submersion sample holder which was filled with 0.96\%-wt. phosphate buffered saline solution (Sigma-Aldrich, St Louis, MO, USA) so that the liquid level was flush with the sample. Neither the saline solution nor the liver samples were degassed in the DMA measurements. A thermocouple was placed in the saline solution so that the temperature of the liver could be monitored in real-time throughout the measurement in the enclosed measurement chamber of the DMA system. A compression clamp was used in all the dynamic modulus measurements.

The dynamic modulus measurement was conducted by first performing a frequency sweep from 0.1 to 200 Hz at room temperature 21 $\pm$ 1 $^\circ$C. The temperature of the chamber was then heated using the built-in heating system of the DMA to 35 $\pm$ 1 $^\circ$C and kept isothermal for 5 minutes to stabilise the temperature of the liver sample. Once the temperature of the liver sample had stabilized, the frequency sweep was performed again. This process was then repeated by increasing the temperature of the sample by 15 $^\circ$C steps until the final temperature of 80 $\pm$ 1 $^\circ$C was reached. The whole measurement process was repeated for 5 different liver samples from different locations of the same liver.

\subsection*{Fitting dynamic modulus to fractional Zener model}

The conventional Zener model of viscoelasticity consists of an elastic (spring) and viscous (damper) elements in series both of which are also parallel with another elastic element. In the fractional Zener model the integer-order time variables of stress and strain are replaced with fractional order variables which allows for more complex responses to be modelled.

In the frequency domain the dynamic modulus is defined by the equation:
\begin{equation}
M^{\ast}(\omega) = M_{\mathrm{S}}(\omega) + \mathrm{i}M_{\mathrm{L}}(\omega)
\end{equation}
where $M_{\mathrm{S}}$ (Pa) is the storage modulus, $M_{\mathrm{L}}$ (Pa) is the loss modulus, $\omega$ (rad$\cdot$s$^{-1}$) is the angular frequency and $\mathrm{i}$ is the imaginary unit. In the Zener model the storage and loss moduli are defined in terms of a relaxation time $\tau$ (s), the spring constant $M_{0}$ (Pa) and the high-frequency limit of the storage modulus $M_{\infty}$ (Pa) by the equations \citep{kohandel2005frequency}:
\begin{align}
\label{eq:zener_storage}
M_{\mathrm{S}}(\omega) &= M_{0}\frac{1+(d+1)\cos(\alpha\pi/2)\omega_{\mathrm{n}}^{\alpha}+d\omega_{\mathrm{n}}^{2\alpha}}{1+2\cos(\alpha\pi/2)\omega_{\mathrm{n}}^{\alpha}+\omega_{\mathrm{n}}^{2\alpha}}
\\
\label{eq:zener_loss}
M_{\mathrm{L}}(\omega) &= M_{0}\frac{(d-1)\sin(\alpha\pi/2)\omega_{\mathrm{n}}^{\alpha}}{1+2\cos(\alpha\pi/2)\omega_{\mathrm{n}}^{\alpha}+\omega_{\mathrm{n}}^{2\alpha}}
\end{align}
where $d = M_{\infty}/M_{0}$, $\omega_{\mathrm{n}} = \omega\tau$ is the normalised frequency, $\alpha$ = 1 for the conventional Zener model and 0 $\le$ $\alpha$ $\le$ 1 for the fractional Zener model. 

Both conventional and fractional Zener models were used to derive temperature dependent viscoelastic parameters from the experimentally measured dynamic modulus data. The magnitude of the modulus was then averaged over the five samples resulting in one frequency dependent magnitude curve at each measurement temperature. Both the conventional and fractional models were then fit to the measured modulus amplitude using a nonlinear least squares method.

\subsection*{Thermal ablation monitoring using harmonic motion imaging}

HMI is an ultrasound-based elasticity imaging technique \citep{maleke2010vivo}, which uses focused, amplitude-modulated waveforms to create an oscillatory ARF \citep{suomi2015optical}. The oscillatory ARF makes the tissue vibrate at the same frequency (typically 50 Hz), which can then be recorded using a speckle-tracking technique. The advantage of HMI is that the amplitude-modulated ultrasound waveforms can also be used to ablate tissue simultaneously during speckle-tracking. This is achieved by using a separate imaging transducer whose centre frequency is different to that of the therapeutic transducer creating the oscillations and thermal ablation. The ability to track displacements continuously (i.e., without sequencing) during HIFU treatment makes HMI a good candidate for real-time therapy monitoring.

\begin{figure}[!htbp]
\begin{center}
\includegraphics[height=6cm]{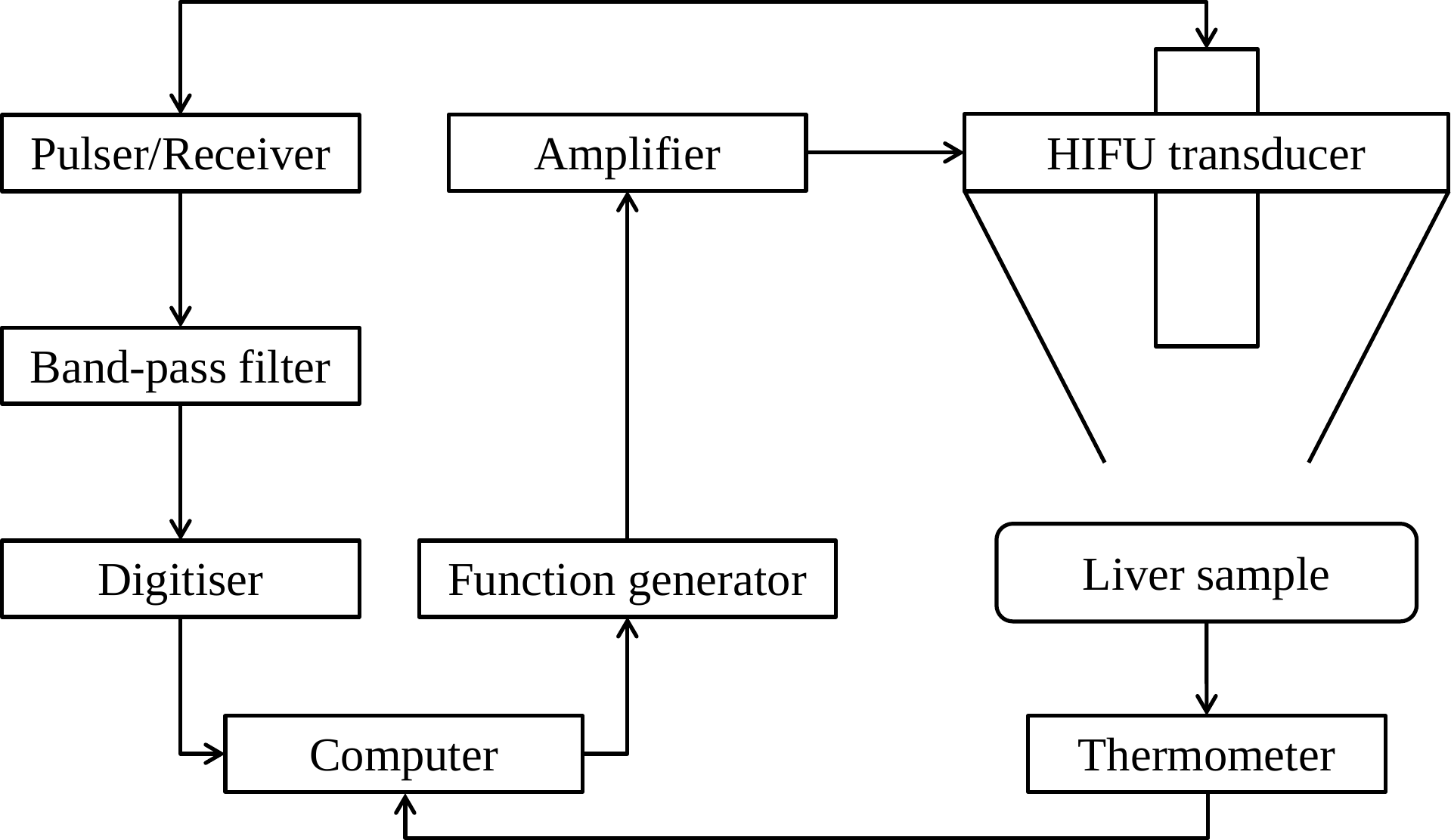}
\end{center}
\caption{Experimental set-up used to acquire displacement data during thermal ablation using harmonic motion imaging (HMI). Thermocouple was placed in the liver sample to record temperature data during sonications.}
\label{fig:measurement_setup}
\end{figure}

The HMI system shown in Figure \ref{fig:measurement_setup} was used to acquire real-time displacement data during thermal ablation in \textit{ex vivo} canine liver. The HMI system was comprised of a 4.755 MHz HIFU transducer (Riverside Research Institute, New York, NY, USA) together with a 7.5 MHz circular single-element transducer (Olympus NDT Inc., Waltham, MA, USA) in the middle. The HIFU transducer had a focal length of 90 mm, an outer diameter of 80 mm and an inner diameter of 16.5 mm (i.e., the diameter of the single-element transducer) and the two transducer were axially aligned. The HIFU transducer was used for ablation as well as to generate ARF for tissue displacements while the inner transducer was used to simultaneously acquire pulse-echo data for displacement tracking. 

A function generator (33522A, Agilent Technologies Inc., Santa Clara, CA, USA) was used to generate a 4.755 MHz signal amplitude-modulated at 50 Hz, which was then amplified (3100L, E\&I, Rochester, NY, USA) for the HIFU transducer. The single-element transducer was operated by a pulser-receiver (5072PR, Olympus, Waltham, MA, USA) with a pulse repetition frequency of 1000 Hz. The received echo signals were transmitted through an analog band-pass filter (Reactel Inc., Gaithersburg, MD, USA), with cut-off frequencies $f_{\textrm{c1}}$ = 5.84 MHz and $f_{\textrm{c2}}$ = 8.66 MHz in order to attenuate HIFU signals, and recorded for data analysis using a high-speed digitiser (Gage Applied, Lockport, IL, USA) with a sampling frequency of 100 MHz.

During the experiments the liver sample with a size of approximately 6 cm $\times$ 6 cm $\times$ 3 cm was placed in a degassed saline solution water bath in room temperature. The temperature acquisition during HIFU thermal ablation was conducted using a thermocouple placed in the middle of the liver sample. The focal point of the HIFU transducer was aligned with the thermocouple by performing raster scans in both axial and lateral planes. The HIFU transducer was moved using a step size of 0.2 mm and at each location of the raster grid a short duration sonication (less than 2 seconds) was performed. The temperature rise due to the short duration sonication was then recorded using a thermometer (HH506RA, Omega Engineering Inc., Stamford, CT, USA) and the raster scans were continued in both planes until the location of the maximum temperature rise was found.

The experimental protocol consisted of multiple sonications in different locations of the liver sample. After the location of the thermocouple was properly aligned with the ultrasound focal point by raster scanning, a 60-second sonication with 7.5 MPa peak-positive pressure was performed using 50 Hz amplitude-modulation. Simultaneously the temperature in the focal point was recorded throughout the whole duration of the sonication and both the displacement and temperature data were saved for data analysis. Once the sonication had finished, the thermocouple was moved to another location so that the lesion created by the previous sonication would not interfere and the whole process was repeated. In post-processing, the acquired peak-to-peak displacement data was normalised to 37 $^{\circ}$C and averaged over 25 samples to suppress the high-frequency noise in the data. A total of 7 sonications were performed of which 4 sets exhibited cavitation which lead to unreliable estimate of the displacement as the cross-correlation algorithm was not able to perform. Therefore data on 3 sonications is reported here.

\subsection*{Thermal ablation simulation model}

The simulations for viscoelastic behaviour of liver during HIFU therapy were conducted in two parts for each temperature step. First, acoustic simulations were conducted to solve the nonlinear ultrasound fields in tissue, from which the temperature field was determined using the bioheat transfer (BHT) equation, and subsequently the size of the lesion. The development of lesion length and diameter during HIFU therapy was estimated using 240 cumulative equivalent minutes at 43 $^{\circ}$C (CEM$_{43^{\circ}\mathrm{C}}$) thermal dose fields from the BHT equation, which was updated to the FEM model at each temperature step. Once these values were obtained, the ultrasound intensity field was converted into ARF field according to (\ref{eq:ARF}) and the force field was then applied to the viscoelastic domain in the FEM model and the tissue deformation was calculated. This iteration was repeated at every temperature step from 35 to 70 $^{\circ}$C at 5 $^{\circ}$C intervals by updating the attenuation, viscoelasticity and lesion size values at each step. This allowed for the evolution of acoustic and viscoelastic parameters and lesion size on tissue deformation to be determined.

\begin{figure*}[!htbp]
\begin{center}
    \subfigure[]
    {
        \includegraphics[height=5cm]{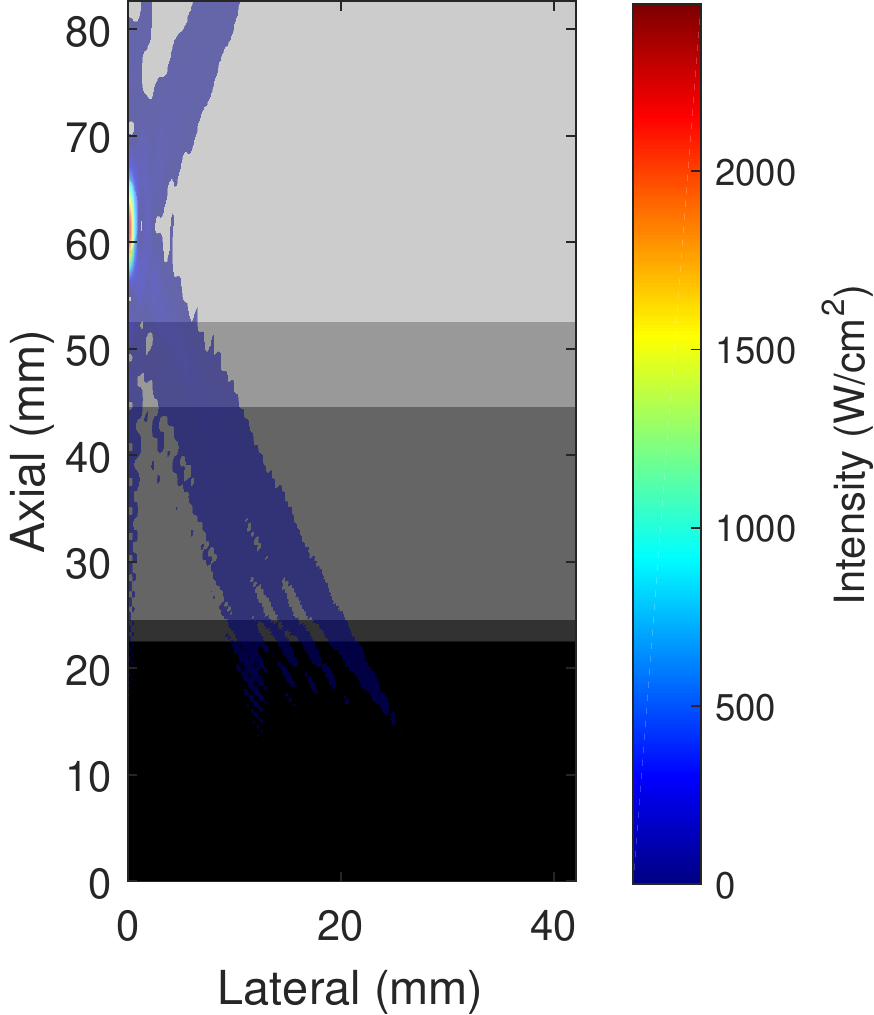}
    }
    \subfigure[]
    {
        \includegraphics[height=5cm]{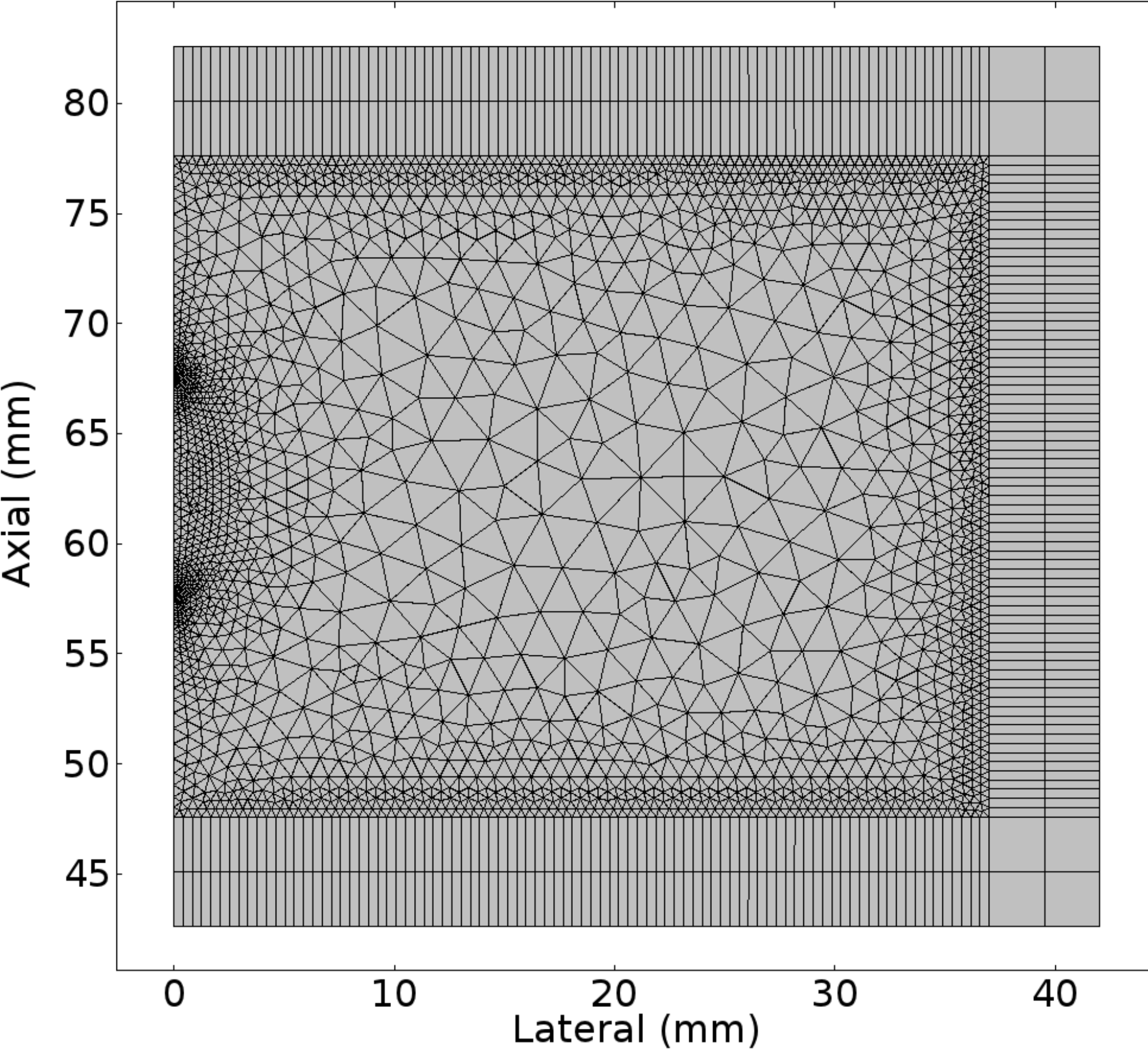}
    }
\end{center}
\caption{(a) 2-D axisymmetric acoustic simulation geometry together with the simulated ultrasound intensity field thresholded at $-$30 dB. The different levels of gray from black to light correspond to layers of water, skin, fat, muscle and liver respectively. (b) 2-D axisymmetric computation mesh used in the finite element method (FEM) model in liver: rectangular elements were used in the infinite element domains and free triangular elements elsewhere.}
\label{fig:FEM_mesh}
\end{figure*}

The simulations of the acoustic fields were conducted in Matlab R2015b (MathWorks Inc, Natick, MA, USA) using a modified version of a HIFU simulator \citep{soneson2009user} which solves the axisymmetric Khokhlov-Zabolotskaya-Kuznetsov (KZK) equation \citep{zabolotskaya1969quasi, kuznetsov1971equations} in the frequency domain using 128 harmonics. The original code was modified to handle multiple tissue layers so that the geometry of the simulations corresponded a real therapeutic situation where the pathway from transducer to liver goes through the layers of water, skin, fat and muscle. The acoustic simulation geometry is presented in Figure \ref{fig:FEM_mesh}(a). Typical values for density, sound speed, attenuation and B/A were used for each tissue type (see Table \ref{tab:acoustic_parameters}), while the polynomial attenuation of the liver was updated at each temperature step. The ultrasound field was calculated as spatial pressure distribution of a HIFU transducer (outer diameter = 6.4 cm, inner diameter = 2.0 cm, focal length = 6.26 cm, frequency = 1.1 MHz). The pressure field of the transducer was solved in the axisymmetric geometry (i.e., half domain) using 2019 grid points in the axial and 672 grid points in the lateral directions. The corresponding ARF field was then calculated using (\ref{eq:ARF}) for each mesh point taking into account the frequency dependent attenuation and applied as point body loads to the spatial domain of the axisymmetric FEM model. The acoustic field calculation was then repeated at each temperature step.

\begin{table}[htbp!]
  \centering
  \caption{Acoustic simulation parameters for different tissue types at 35 $^{\circ}$C. $^{(*)}$ The attenuation of liver was updated at each temperature step according to the fourth order polynomial dependence.}
    \begin{tabular}{lccccc}
    \hline
          		& Density   		& Sound speed   & Attenuation 				& n		& B/A 	\\
          		& (kg/m$^{3}$) 		& (m/s) 		& (dB/MHz$^{\mathrm{n}}$/cm)& 		& 		\\
    \hline
    Water 		& 1000  			& 1482  		& 0.00217 					& 2.00	& 5.0 	\\
    Skin  		& 1090  			& 1615  		& 0.35    					& 1.55	& 7.9 	\\
    Fat   		& 950   			& 1478  		& 0.48  					& 1.20	& 10.0 	\\
    Muscle		& 1050  			& 1547  		& 1.09   					& 1.10	& 7.4 	\\
    Liver$^{*}$ & 1060  			& 1584			& 0.35				  		& 1.00	& 7.2 	\\
    \hline
    \end{tabular}
  \label{tab:acoustic_parameters}
\end{table}

A FEM model (Comsol Multiphysics 4.4, Comsol Inc, Burlington, MA, USA) was used to simulate the ARF induced displacements. In the FEM model a 2-D axisymmetric geometry for the liver was used so that the symmetry axis was set along the axis of the HIFU transducer (see Figures \ref{fig:FEM_mesh}(b) and (c)). The liver had a size of 42 mm (lateral) $\times$ 40 mm (axial) corresponding to the size of the liver in the acoustic simulations (i.e., the calculated ARF field size). Furthermore, 5 mm infinite element layers on the outer edges were used in order to minimise reflections from the boundaries. The outer surface of the liver was constrained to reflect a situation where the liver is in a fixed position within the body. The mechanical properties of the liver were modelled as a linear, isotropic and viscoelastic solid using properties derived from measurements and are defined later in the results section. A free triangular mesh was utilised in the central area of the liver and a rectangular mesh in the infinite element layers. Using triangular elements for nearly incompressible materials can lead to problems with ill-conditioning, which was overcome by utilizing mixed formulation in Comsol. A convergence study was run before the actual simulations to determine the proper element size for the computation mesh.

\subsection*{Simulation execution}

The ARF induced displacements were calculated for a 60-second sonication resulting in a temperature range 35-70 $^{\circ}$C with 5 $^{\circ}$C intervals. At each discrete temperature point during the sonication both the viscoelastic Zener model parameters and the attenuation of the liver and the lesion were set to correspond to the measured values. The attenuation power law coefficient of the liver was kept at a constant value of 1.0 because no significant changes had been observed in the literature. The liver tissue was then pushed using a rectangular 0.5 ms pulse and the dynamic response was tracked at the geometric focus of the transducer. The displacement value at 0.5 ms was recorded and the viscoelastic and attenuation parameters were set for the next temperature iteration. It should be noted that the lesion did not appear until 60 $^{\circ}$C was achieved, and hence, only the attenuation changed up to this point. At temperatures of 60 $^{\circ}$C and above the viscoelastic parameters and attenuation were adjusted within the lesion while the parameters of the surrounding liver tissue were kept constant. For the lesion area the viscoelastic properties varied linearly from the peak temperature at the centre to the background values at the edge in order to reduce shear wave reflections.

In the second simulation study a Monte-Carlo style approach was conducted to determine the effect of variation in tissue property changes. This was done by varying the spring constant $M_{\infty}-M_{0}$ (i.e., the elastic branch in the Zener model) of the liver between 5-155 kPa and the attenuation between 0.15-1.35 dB/cm which are the typical ranges before and after thermal ablation found in the literature (recall Figure \ref{fig:literature_att}(b) and Figure \ref{fig:literature_ym}). Varying the values of $M_{0}$ and $\tau$ within the range of their variance in the experimental data did not result in changes in the displacement amplitude due to the overpowering effect of $M_{\infty}$. Therefore, the Zener model viscoelastic branch parameters were kept constant at $M_{0}$ = 8.0 Pa and $\tau$ = 0.5 s representing the average values of liver tissue in the experiments. In the simulation model the values were changed for the whole area of the liver phantom (i.e., no lesion present) which is similar to the situation with one large or multiple lesions next to each other. At each elastic modulus/attenuation coefficient combination the tissue was pushed with a 0.5 ms long pulse and the displacement amplitude was recorded at the geometric focus of the transducer.

\pagebreak

%%%%%%%%%%% Results
\section*{Results}

\subsection*{Temperature dependence of viscoelastic properties in liver}

\begin{figure*}[!htbp]
\begin{center}
    \subfigure[]
    {
        \includegraphics[width=0.8\textwidth]{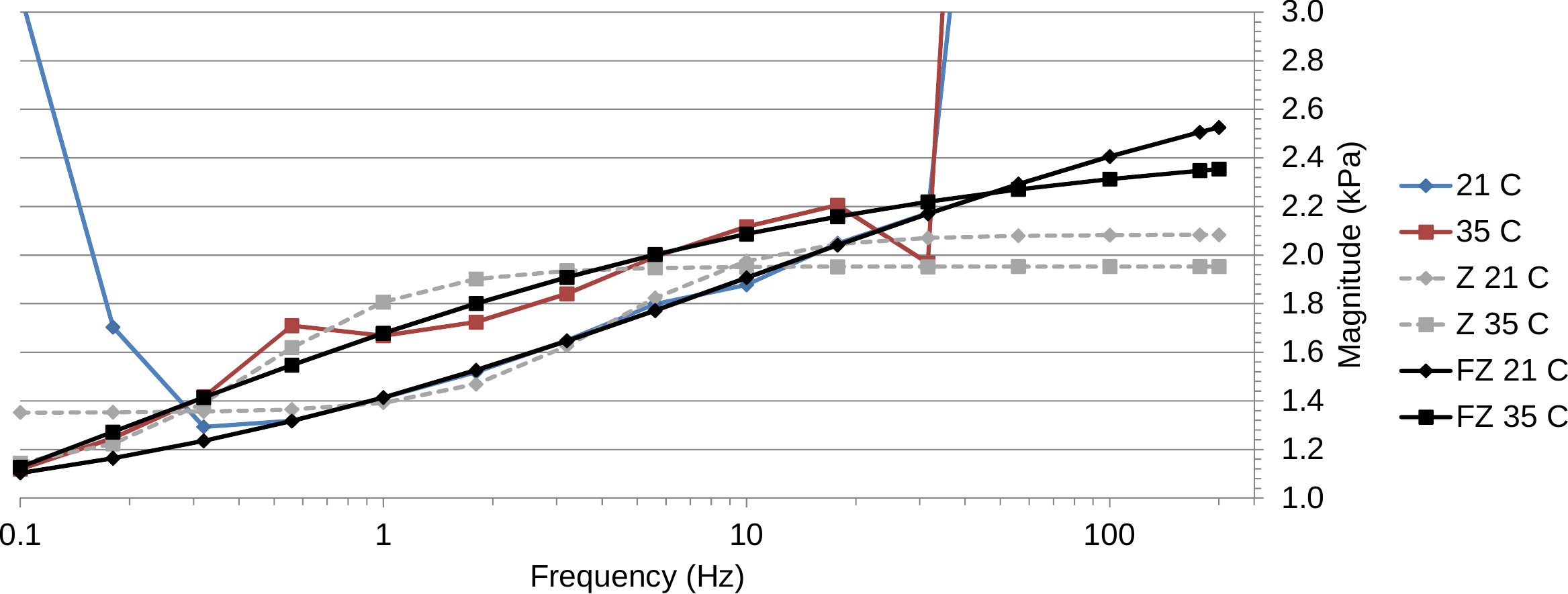}
    }
    \\
    \subfigure[]
    {
        \includegraphics[width=0.8\textwidth]{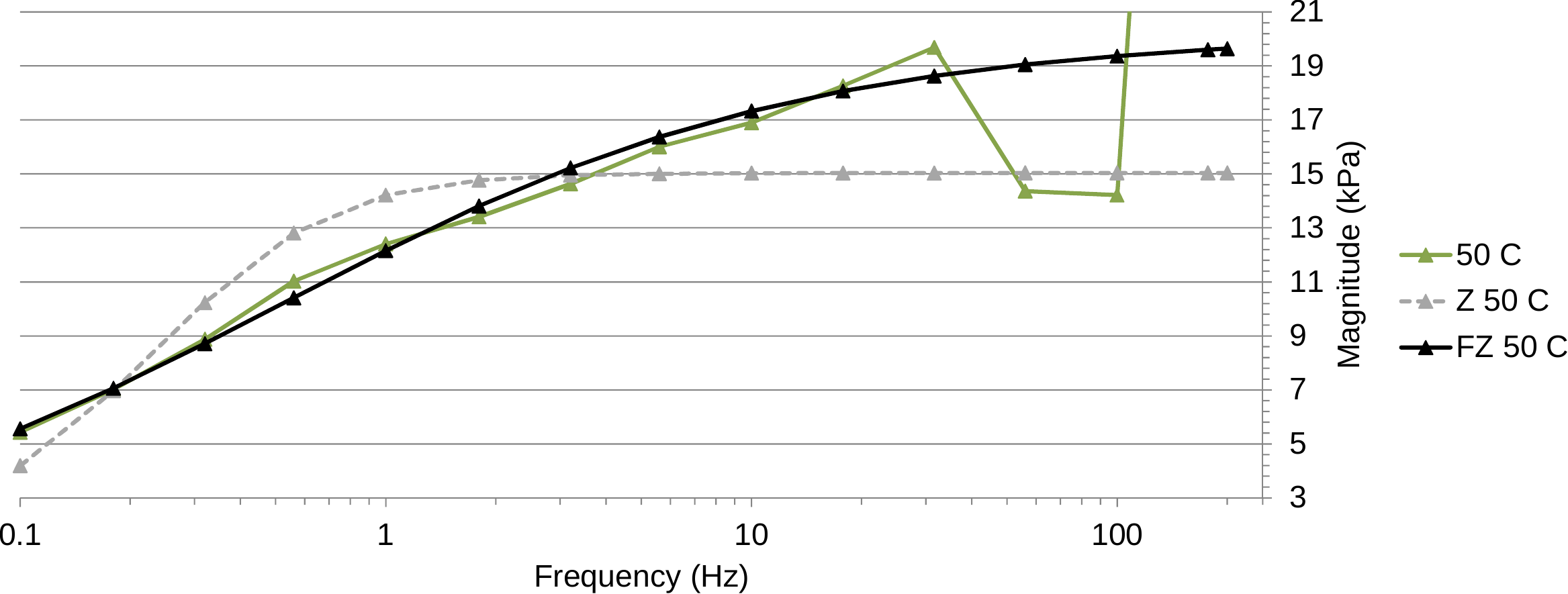}
    }
    \\
    \subfigure[]
    {
        \includegraphics[width=0.8\textwidth]{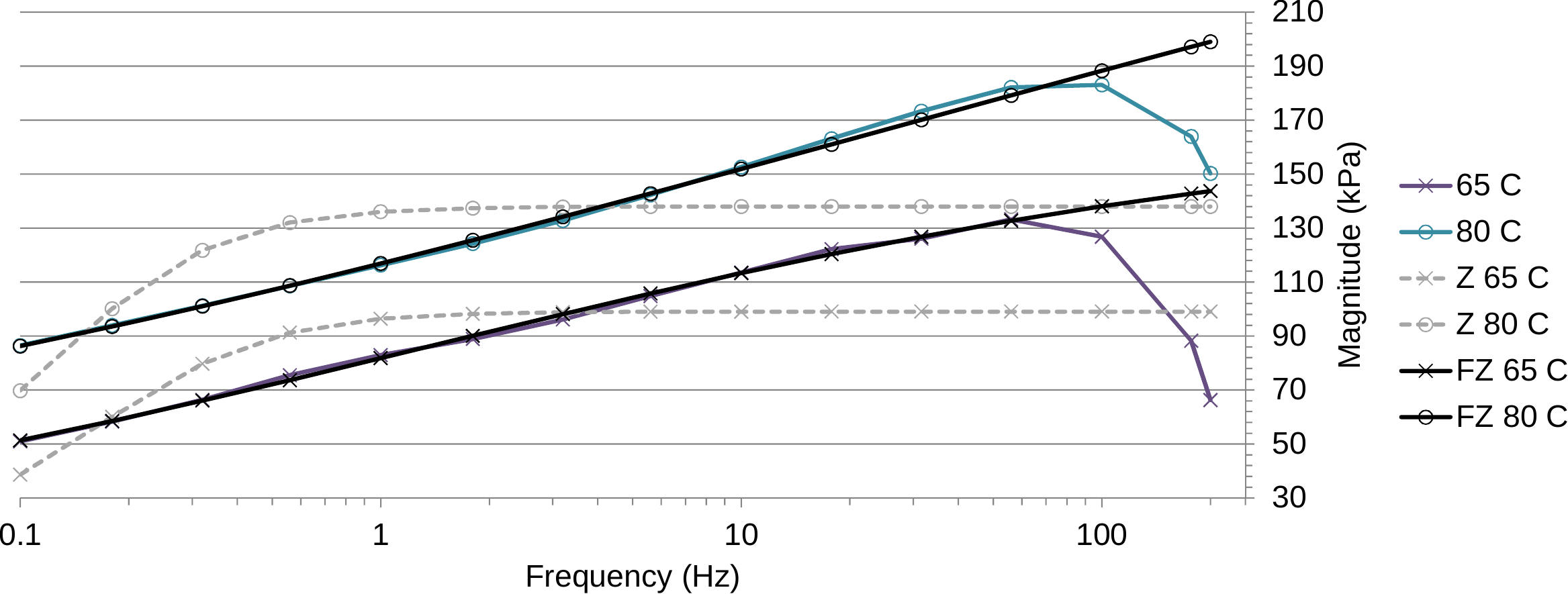}
    }
\end{center}
\caption{The average (5 samples) magnitude of the compressional dynamic modulus of \textit{ex vivo} bovine liver samples during heating at temperatures of (a) 21 to 35 $^{\circ}$C, (b) 50 $^{\circ}$C and (c) 65 to 80 $^{\circ}$C over the frequency range from 0.1 to 200 Hz. Conventional and fractional Zener models of viscoelasticity were fit to the data.}
\label{fig:dynamic_modulus}
\end{figure*}

\begin{figure*}[!htbp]
\begin{center}
    \subfigure[]
    {
        \includegraphics[width=0.8\textwidth]{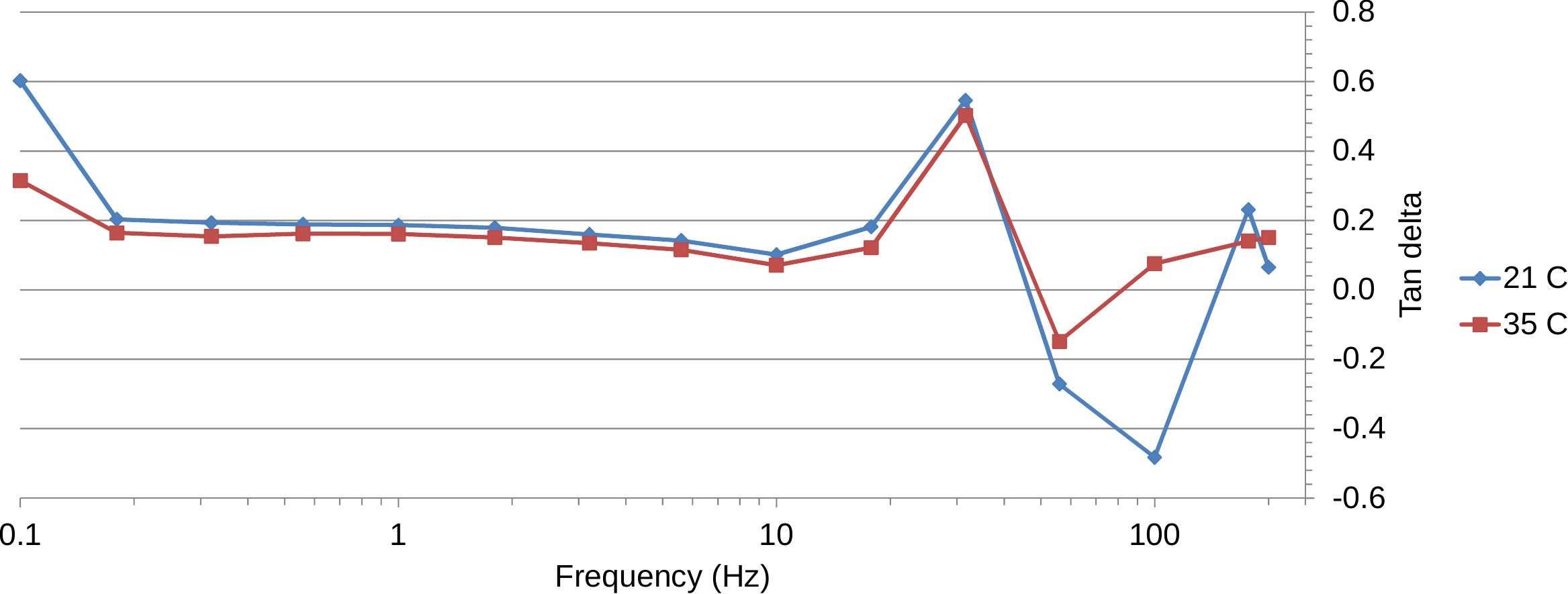}
    }
    \\
    \subfigure[]
    {
        \includegraphics[width=0.8\textwidth]{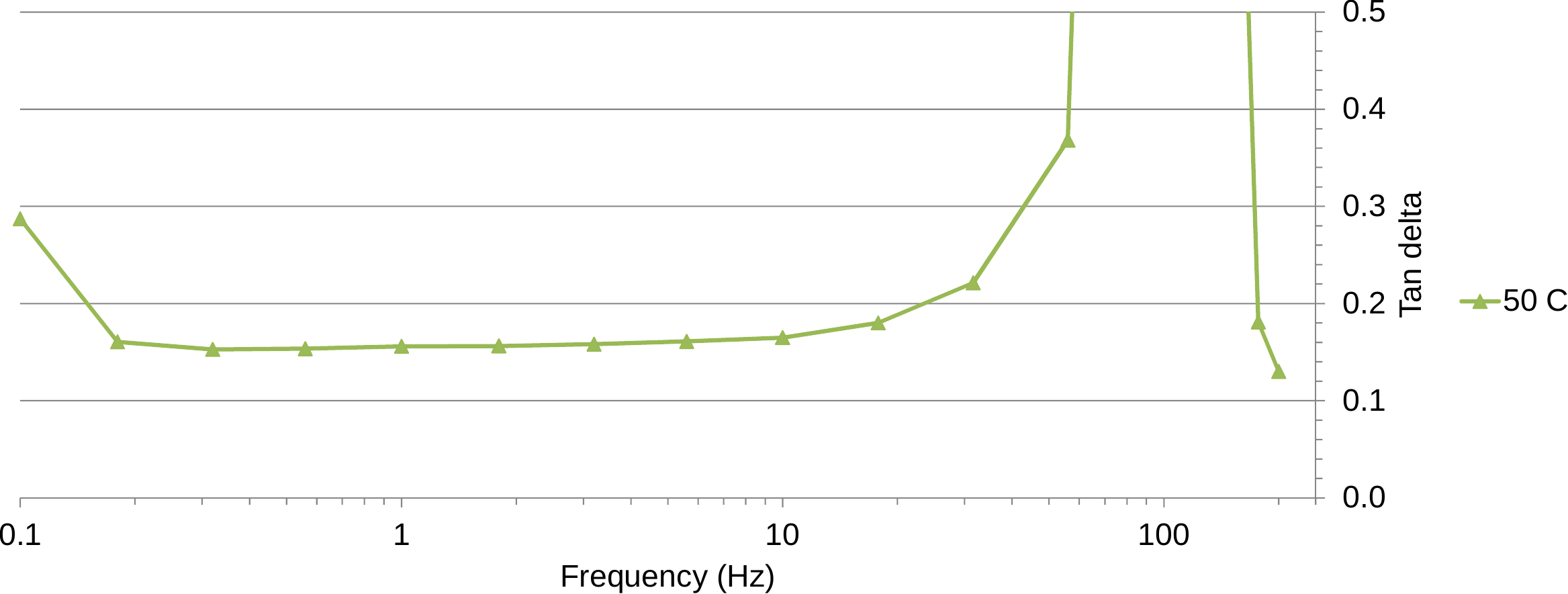}
    }
    \\
    \subfigure[]
    {
        \includegraphics[width=0.8\textwidth]{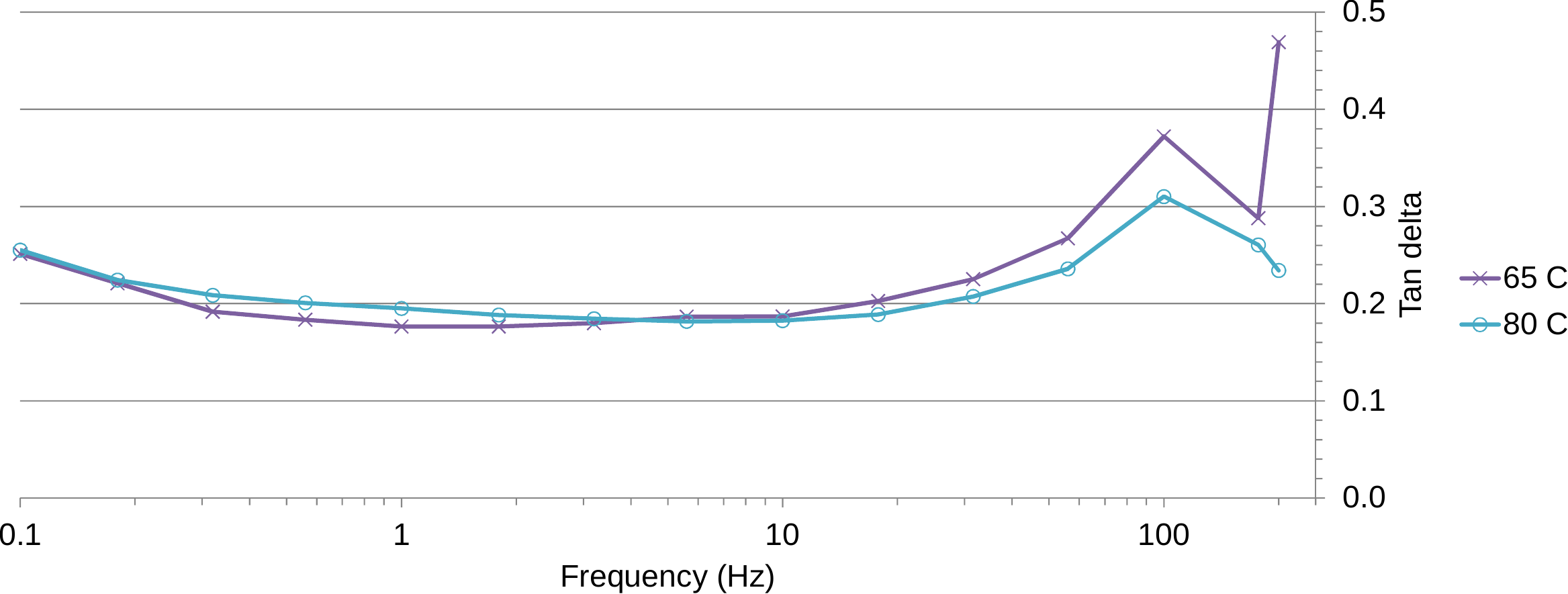}
    }
\end{center}
\caption{The average (5 samples) $\tan \delta$ phase values of \textit{ex vivo} bovine liver samples during heating at temperatures of (a) 21 to 35 $^{\circ}$C, (b) 50 $^{\circ}$C and (c) 65 to 80 $^{\circ}$C over the frequency range from 0.1 to 200 Hz.}
\label{fig:tan_delta}
\end{figure*}

The experimentally measured magnitude of the dynamic modulus averaged over the five of samples is presented in Figure \ref{fig:dynamic_modulus} over the frequency range 0.1-200 Hz at temperatures from 21 to 80 $^{\circ}$C. The corresponding phase values (i.e., $\tan \delta$) are presented in Figure \ref{fig:tan_delta}. For both conventional and fractional Zener models the magnitude of the average modulus were fit by using a nonlinear least squares method. The viscoelastic fitting parameters are shown in Table \ref{tab:zener_parameters} for both models. The frequency ranges for the fits were chosen to exclude the rapidly fluctuating or increasing magnitude values which the models cannot capture. The high stiffness below 0.3 Hz at 21 $^{\circ}$C is most likely caused by low signal to noise ratio at these frequencies. At high frequencies (above 30-40 Hz), rapid fluctuations in the magnitude of dynamic modulus were observed at all temperatures. This behaviour was also noted by \citet{kiss2004viscoelastic} who postulated them to be an electronic signal-to-noise ratio artefact in the force measurement. It is likely that at high frequencies the DMA system cannot distinguish between ambient system noise and the response due to the tissue. This effect can also be seen in Figure \ref{fig:tan_delta} where the phase is rapidly fluctuating to due to the presence of experimental artefacts at high frequencies.

At 21 $^{\circ}$C and 35 $^{\circ}$C temperatures in Figure \ref{fig:dynamic_modulus}(a) the conventional Zener model predicts the magnitude to increase from a initial value of 1.2-1.4 kPa up to approximately 2.0 kPa for both temperatures. The modulus is predicted to stabilise at frequencies above 10 Hz. The fractional model predicts the magnitude to increase from a initial value of around 1.2 kPa to approximately 2.4 kPa at 200 Hz for both temperatures. 

At 50 $^{\circ}$C temperature in Figure \ref{fig:dynamic_modulus}(b) the conventional Zener model predicts the magnitude to increase from a initial value of 5 kPa up to 15 kPa stabilising at frequencies above 1 Hz. The fractional model follows the experimental data better at higher frequencies reaching approximately 19 kPa at 200 Hz.

At 65 $^{\circ}$C in Figure \ref{fig:dynamic_modulus}(c) the conventional Zener model starts from the initial value around 40 kPa stabilising to approximately 100 kPa at frequencies above 1 Hz. At 80 $^{\circ}$C the values increase from around 70 kPa up to 140 kPa. The fractional model again follows the experimental data better reaching values of approximately 140 kPa and 200 kPa for 65 $^{\circ}$C and 80 $^{\circ}$C temperatures respectively.
 
In general the sum of squared errors (SSE) (see Table \ref{tab:zener_parameters}) of the fractional model are two magnitudes lower compared to the conventional model for all temperatures. This is because the fractional model better captures the continuous increase in the magnitude and is not trying to stabilise the values at high frequencies.

\begin{table}[htbp]
  \centering
  \caption{Conventional and fractional Zener model fitting parameters for thermally ablated liver at different temperatures. The sum of squared errors (SSE) is also shown in each case.}
    \begin{tabular}{lrrrrrr}
    \hline
    	& \multicolumn{1}{c}{$f$ (Hz)} & \multicolumn{1}{c}{$M_{\infty}$ (kPa)} & \multicolumn{1}{c}{$M_{0}$ (Pa)} & \multicolumn{1}{c}{$\tau$ (s)} & \multicolumn{1}{c}{$\alpha$} & \multicolumn{1}{c}{SSE} \\
    \hline
    Z 21 $^{\circ}$C & 0.6-31.6 & 2.1   & 1351.1 & 3.45E$-$02 & 1 & 0.008 \\
    Z 35 $^{\circ}$C & 0.1-17.8 & 2.0   & 1102.3 & 3.09E$-$01 & 1 & 0.044 \\
    Z 50 $^{\circ}$C & 0.1-31.6 & 15.0  & 8.0   & 4.62E$-$01 & 1 & 0.237 \\
    Z 65 $^{\circ}$C & 0.1-56.0 & 99.1  & 8.1   & 6.74E$-$01 & 1 & 0.349 \\
    Z 80 $^{\circ}$C & 0.1-100.0 & 138.0 & 8.1   & 9.32E$-$01 & 1 & 0.366 \\
    \hline
    FZ 21 $^{\circ}$C & 0.6-31.6 & 3.0   & 855.2 & 1.23E$-$02 & 0.40 & 0.000 \\
    FZ 35 $^{\circ}$C & 0.1-17.8 & 2.5   & 8.0   & 7.52E$-$01 & 0.38 & 0.014 \\
    FZ 50 $^{\circ}$C & 0.1-31.6 & 20.2  & 8.1   & 2.24E$-$01 & 0.55 & 0.011 \\
    FZ 65 $^{\circ}$C & 0.1-56.0 & 171.2 & 8.1   & 9.99E$-$02 & 0.32 & 0.002 \\
    FZ 80 $^{\circ}$C & 0.1-100.0 & 335.5 & 8.1   & 5.20E$-$03 & 0.19 & 0.002 \\
    \hline
    \end{tabular}
  \label{tab:zener_parameters}
\end{table}

\pagebreak

\subsection*{Experimental displacements during thermal ablation}

\begin{figure}[!htbp]
\begin{center}
\includegraphics[width=1\textwidth]{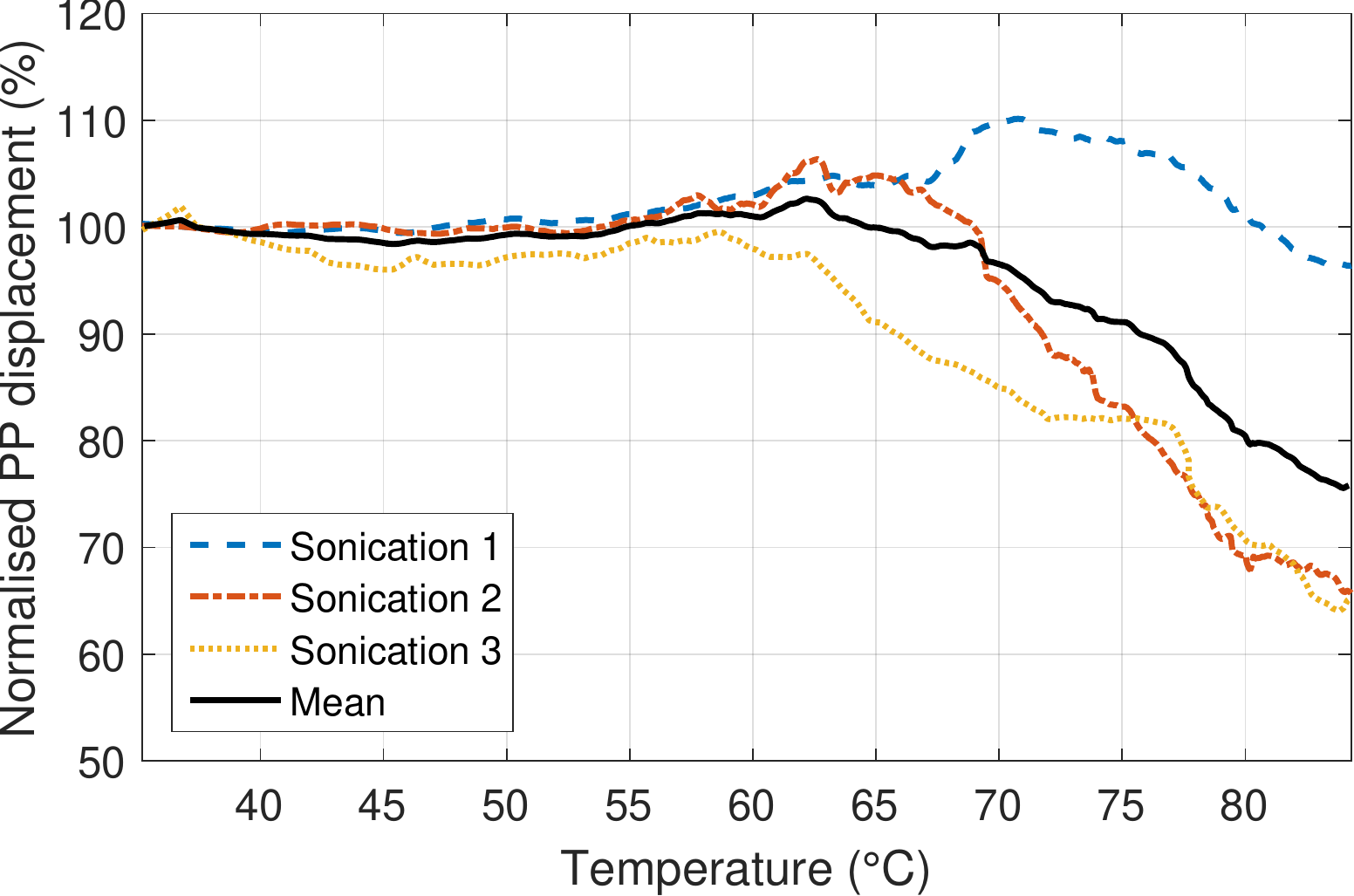}
\end{center}
\caption{Experimental HMI data showing the change in normalised (to 37 $^{\circ}$C) peak-to-peak displacement amplitude with temperature during three different 60-second sonications in \textit{ex vivo} canine liver.}
\label{fig:pp_disp_with_temperature}
\end{figure}

The HMI system and thermocouple were used to monitor the evolution of the ARF induced displacements over 60-second sonications in \textit{ex vivo} liver. The normalised (to 37 $^{\circ}$C) and averaged peak-to-peak displacement data as a function of temperature for three different sonications are presented in Figure \ref{fig:pp_disp_with_temperature}. The mean for the same data is also shown. For the first sonication the peak-to-peak amplitude decreases to a minimum value of 99\% at approximately 45-50 $^{\circ}$C after which an increase to the peak value of 110\% at around 70 $^{\circ}$C is observed. After the peak value there is a relative slow decrease to 96\% at the temperature of 84 $^{\circ}$C. Similar behaviour is observed in the case of the second sonication where the peak-to-peak displacement decreases to 99\% at around 45 $^{\circ}$C and the increases reaching a peak value of 105\% at approximately 60-65 $^{\circ}$C. After the peak value the curve decreases to 65\% at 84 $^{\circ}$C. In the third sonication the amplitude first decreases to 96\% around 45 $^{\circ}$C after which a peak value of 100\% at 59 $^{\circ}$C is reached. After the peak the minimum value of 65\% at the temperature of 84 $^{\circ}$C is measured.

It should be noted that the peak displacements occur at different temperatures in different sonications. For example at around 70 $^{\circ}$C the amplitude has risen approximately to 110\% in the first sonication whereas in the second and third sonications the amplitudes have decreased to 95\% and 85\% respectively. This suggest that the time history of the displacements during HIFU therapy is an important indicator of the state of the ablation as the decrease in amplitude does not occur at a constant temperature. This effect can also been seen in the large deviation of the displacement estimates after about 60 $^{\circ}$C where the stiffness changes due to thermal ablation has been shown to occur \citep{wu2001assessment, sapin2010temperature}.

\pagebreak

\subsection*{Simulations of thermal ablation}

The simulations of ARF induced displacements during HIFU employed the temperature dependent models described in the simulation model section with the polynomial constants for attenuation in Figure \ref{fig:literature_att_norm} and the viscoelastic parameters in Table \ref{tab:FEM_parameters}. The Poisson's ratio for both liver and lesion was set to 0.499.

The viscoelastic parameters for liver tissue were kept constant at all temperatures while the parameters for lesion were changed at 60 $^{\circ}$C and above when the lesion started to form. In addition to the average values presented in the tables, the attenuation and elastic modulus $M_{\infty}-M_{0}$ of the lesion were individually changed $\pm$25\% to examine their effect on the displacements. The effect of temperature dependent sound speed and attenuation power law coefficient were also studied by changing the corresponding lesion parameters by $\pm$0.5\% and $\pm$25\%, respectively. However, no effect on the displacement was observed due to high stiffness of the lesion at various temperatures, and hence, their temperature dependence was excluded from the final simulations.

\begin{table}[!htbp]
  \centering
  \caption{Temperature dependent viscoelastic parameters for background liver tissue and lesion used in the finite element method (FEM) model.}
    \begin{tabular}{lrrrr}
    \hline
    	& Liver 	& \multicolumn{3}{c}{Lesion} \\
    \hline
    $T$ ($^{\circ}$C) & 35-70 & 60	& 65	& 70	\\
    $M_{0}$ (Pa) &   1.1   & 8.1 & 8.1 & 8.1   \\
    $M_{\infty}-M_{0}$ (kPa) & 0.9 & 34.7  & 52.6  & 79.6  \\

    $\tau$ (s) &   0.309  & 0.626 & 0.699 & 0.772  \\
    Lesion length (mm) &   \multicolumn{1}{c}{N/A}  & 3.0   & 5.0   & 6.8   \\
    Lesion diameter (mm) &  \multicolumn{1}{c}{N/A}  & 0.6   & 1.2   & 1.7   \\
	\hline
    \end{tabular}
  \label{tab:FEM_parameters}
\end{table}

\begin{figure}[!htbp]
\begin{center}
\includegraphics[width=1\textwidth]{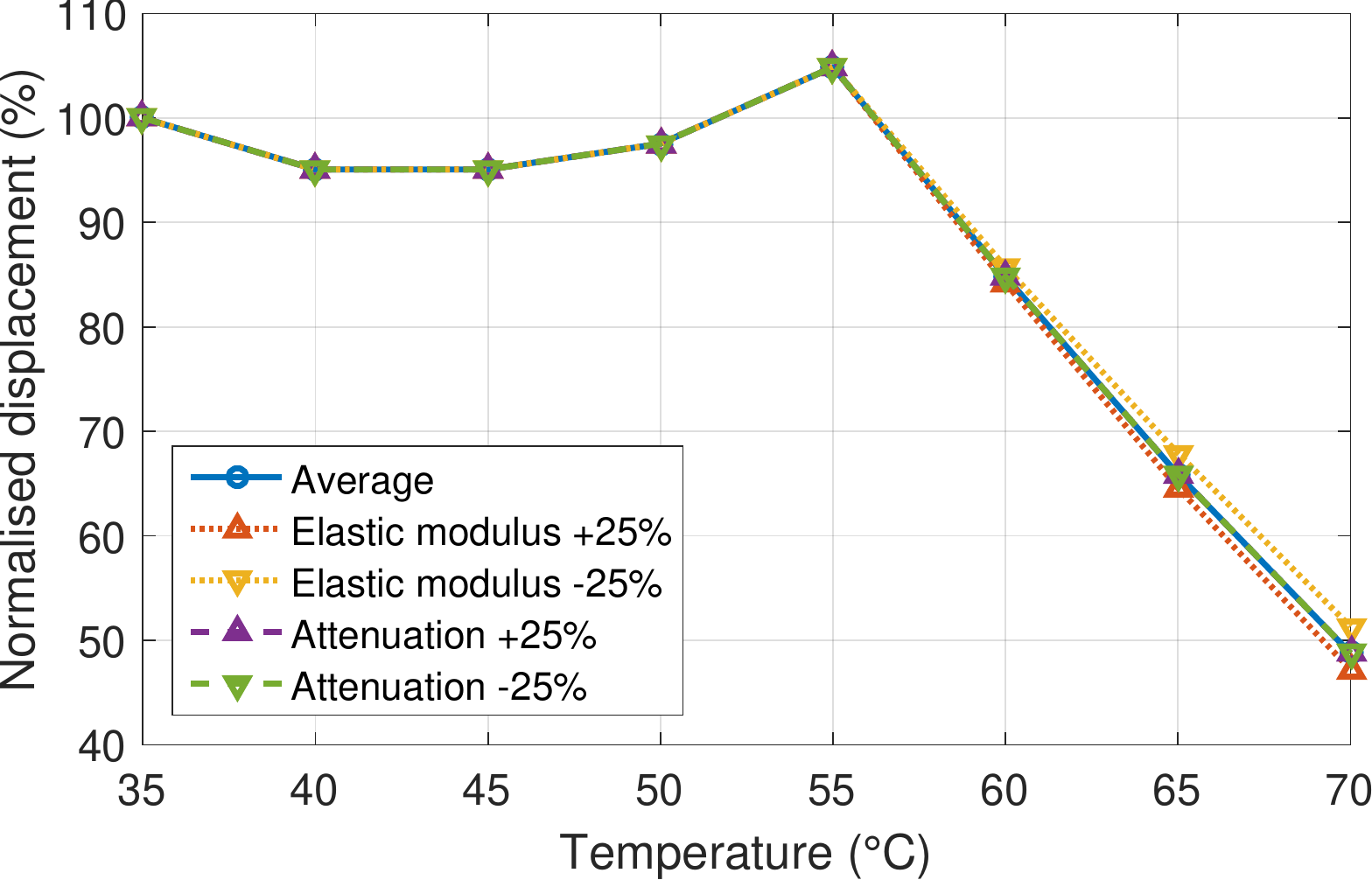}
\end{center}
\caption{Simulated data showing the change in normalised (to 35 $^{\circ}$C) ARF displacement amplitude with temperature during HIFU therapy in liver using the parameters from Table \ref{tab:FEM_parameters}. The values for attenuation and elastic modulus of the lesion were changed $\pm$25\% to examine their effect with respect to the average response.}
\label{fig:pp_disp_with_temperature_simulation}
\end{figure}

The simulated evolution of the normalised (to 35 $^{\circ}$C) displacement amplitude with temperature during HIFU therapy is presented in Figure \ref{fig:pp_disp_with_temperature_simulation}. The average displacement amplitude starts to decrease from the initial value of 100\% at 35 $^{\circ}$C down to 95\% at approximately 40-45 $^{\circ}$C due to the changes in the polynomial attenuation of liver. After the initial decrease the amplitude starts to increase (again due to attenuation) peaking 105\% at 55 $^{\circ}$C. At 60 $^{\circ}$C the amplitude rapidly decreases down to 85\% due to the lesion growth. Although the attenuation of the lesion is still growing at this point, the effect is overshadowed by the large changes in lesion stiffness. At higher temperatures, the lesion continues to grow in size and stiffness which brings the displacement amplitude down to 49\% at the final temperature of 70 $^{\circ}$C in the end of the sonication.

In addition to the average response, the displacement amplitudes are shown by changing the attenuation and elastic modulus of the lesion by $\pm$25\%. Changing attenuation by $\pm$25\% resulted in a negligible change in displacement amplitude (less than 0.1\%). Changing elastic modulus of the lesion by $\pm$25\% resulted in a small change in displacement for temperatures of 60 $^{\circ}$C and above, but differences of approximately 2\%-points were observed at 70 $^{\circ}$C. This suggest that changes in attenuation and displacement have very little effect on the displacement amplitude, and that displacement is robust to these changes.

\begin{figure*}[!htbp]
\begin{center}
    \subfigure[]
    {
        \includegraphics[width=0.3\textwidth]{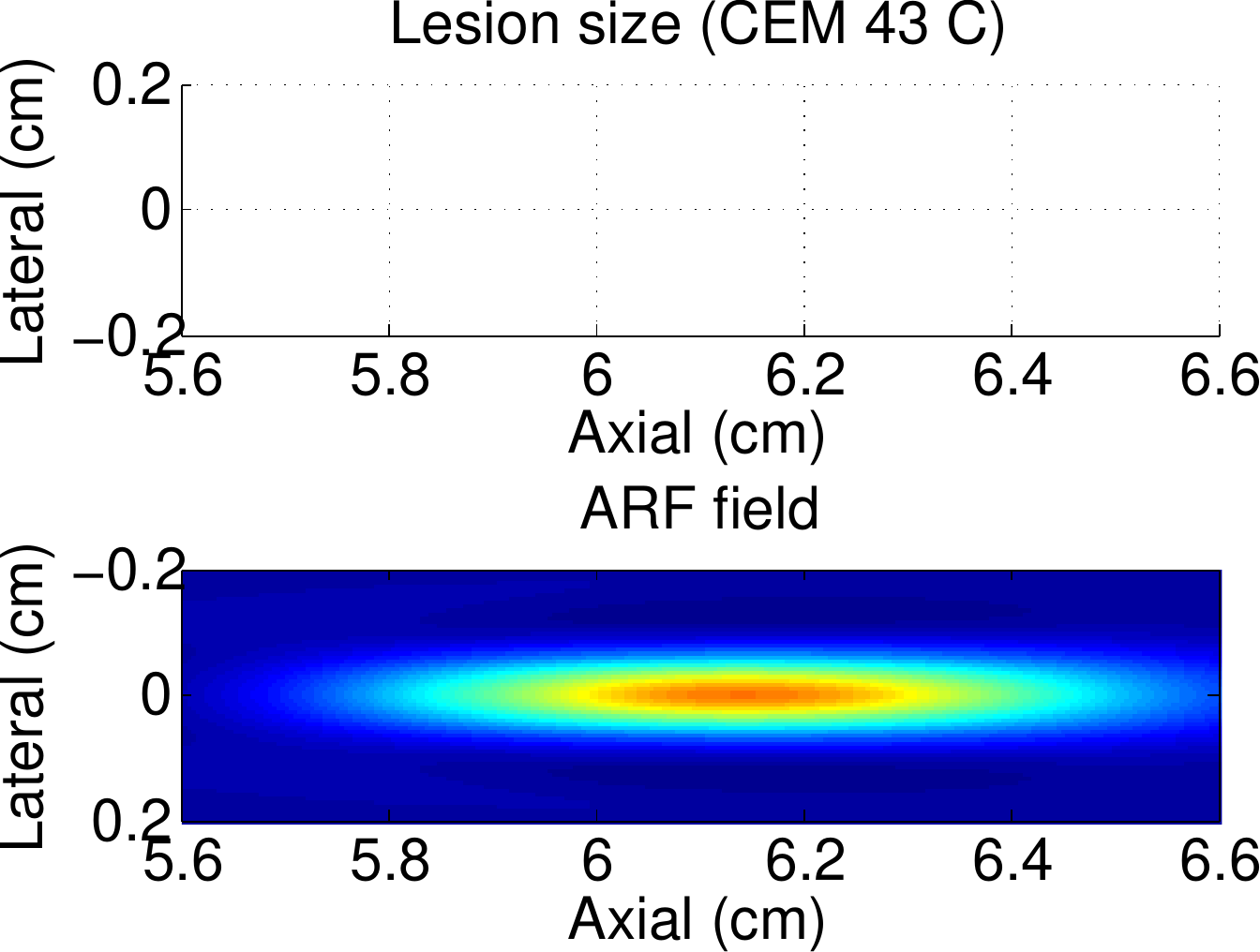}
    }
    \subfigure[]
    {
        \includegraphics[width=0.3\textwidth]{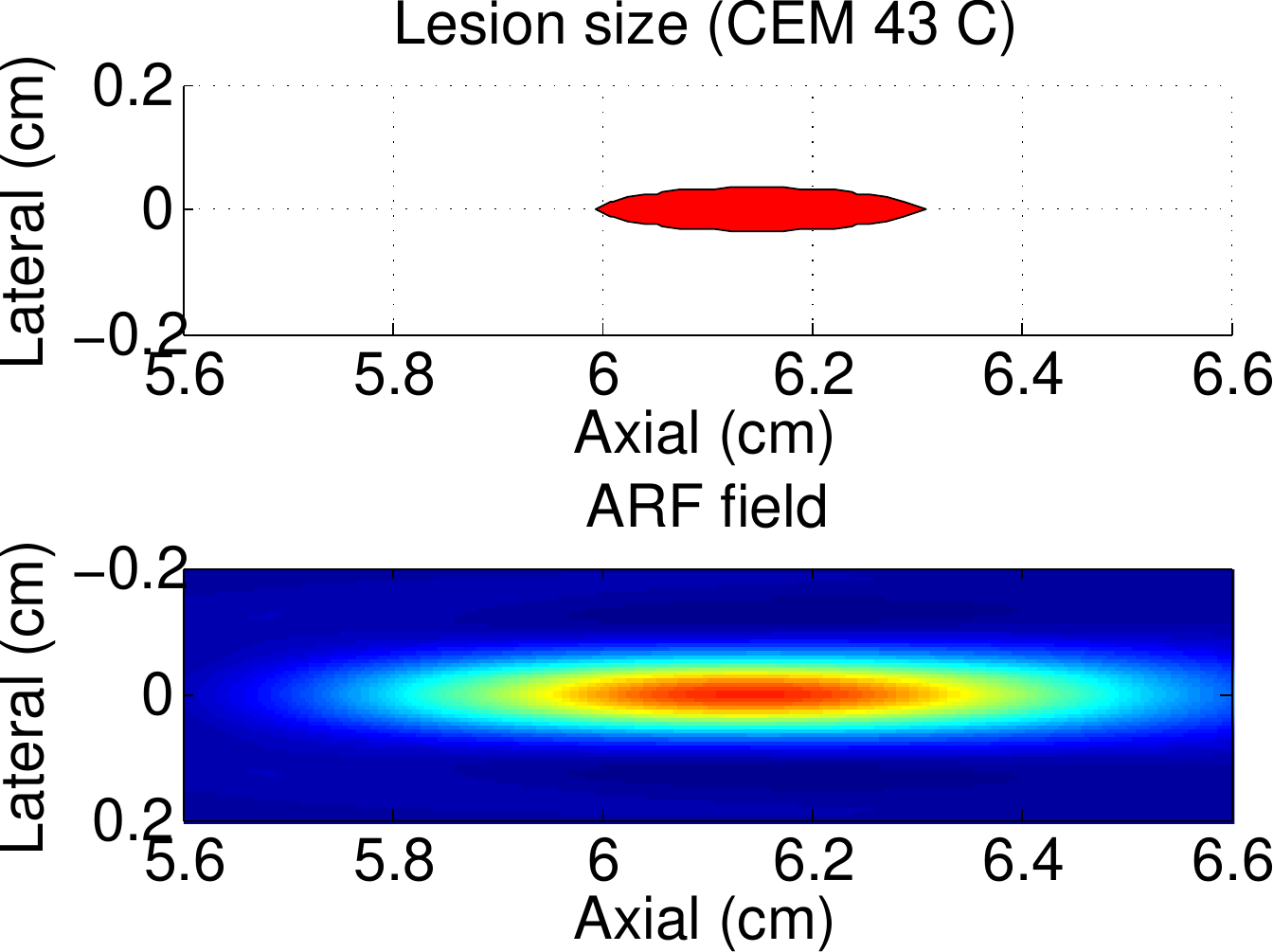}
    }
    \subfigure[]
    {
        \includegraphics[width=0.3\textwidth]{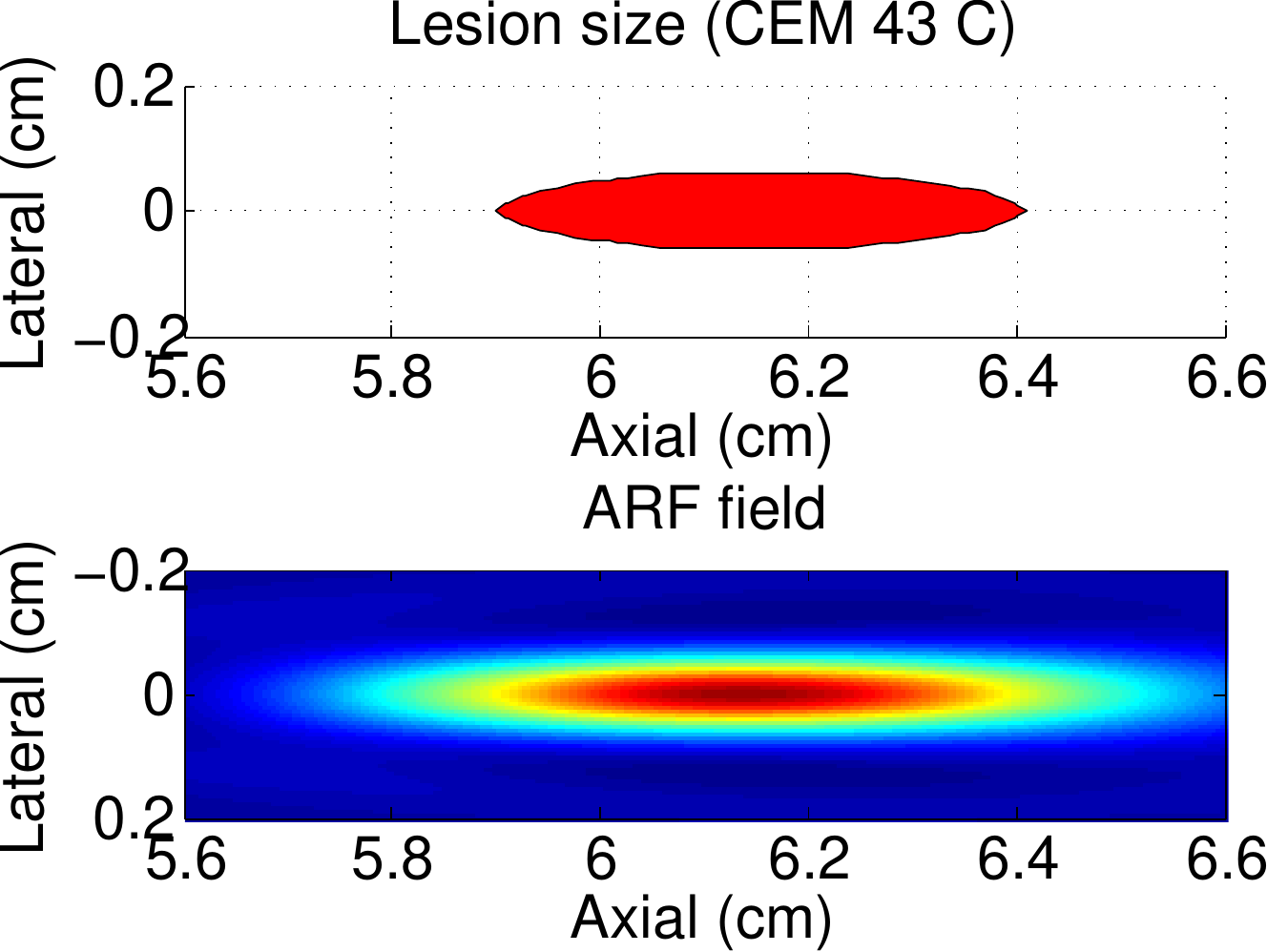}
    }
\end{center}
\caption{The evolution of the lesion size and the ARF field at (a) 55, (b) 60 and (c) 65 $^{\circ}$C, which can be related to changes in displacement in Figure \ref{fig:pp_disp_with_temperature_simulation}.}
\label{fig:lesion_size_vs_ARF}
\end{figure*}

The growth of the lesion size with respect to the ARF field is shown in Figure \ref{fig:lesion_size_vs_ARF}. At 55 $^{\circ}$C (see Figure \ref{fig:lesion_size_vs_ARF}(a)) a lesion has not formed yet but the magnitude of the ARF field has increased due to the polynomial dependence of the attenuation, which causes the displacement amplitude to increase to 105\% in Figure \ref{fig:pp_disp_with_temperature_simulation}. At 60 $^{\circ}$C (see Figure \ref{fig:lesion_size_vs_ARF}(b)) the lesion has grown approximately the size of the ARF field, whose magnitude still keeps increasing, however the displacement decreases in Figure \ref{fig:pp_disp_with_temperature_simulation} due to the stronger effect of stiffness growth compared to the attenuation. At 65 $^{\circ}$C (see Figure \ref{fig:lesion_size_vs_ARF}(c)) the size of the lesion has become larger than the ARF field and the displacement in Figure \ref{fig:pp_disp_with_temperature_simulation} continues to decrease steeply as the increase in stiffness dominates the response.

\begin{figure}[!htbp]
\begin{center}
\includegraphics[width=1\textwidth]{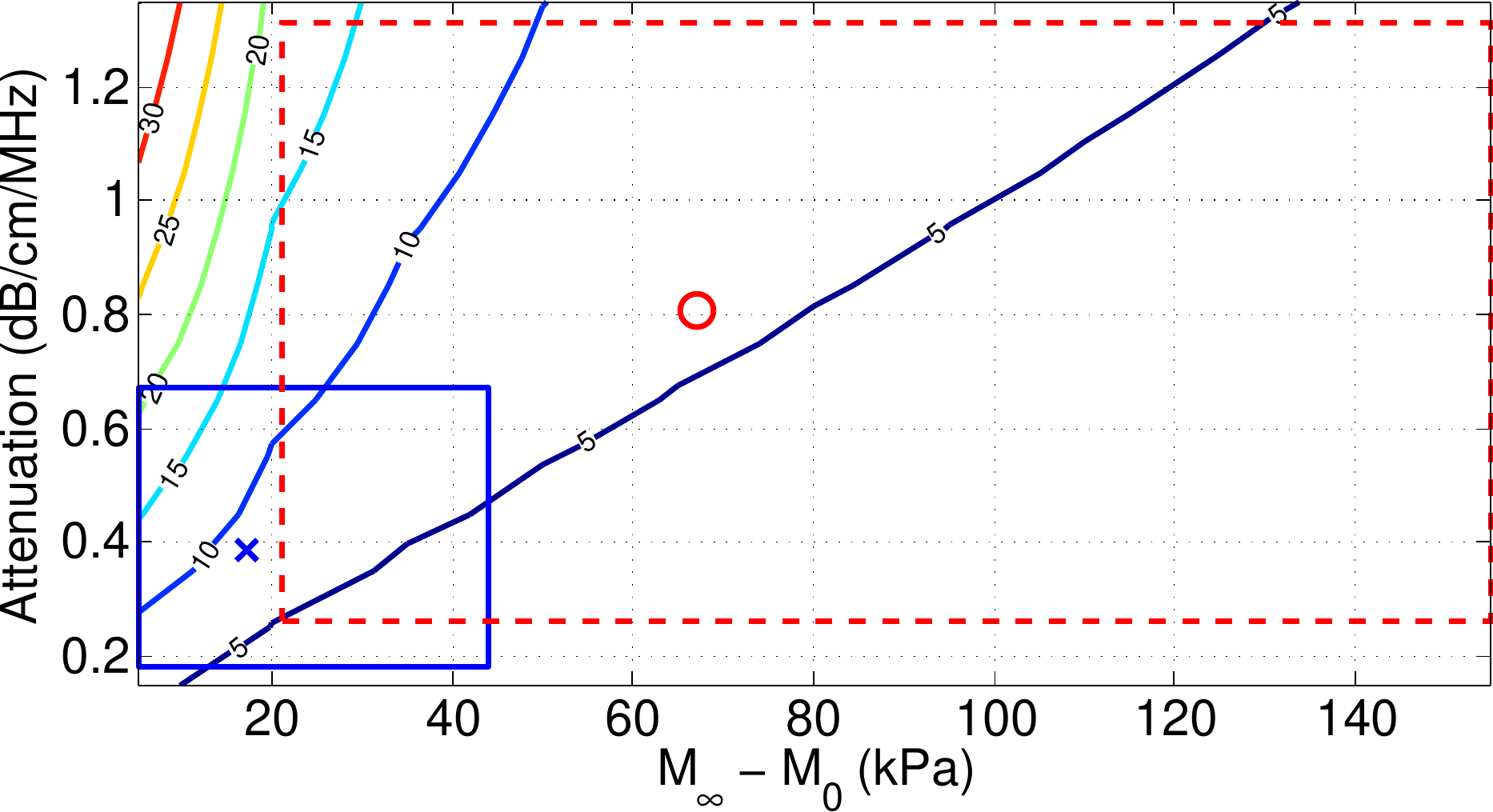}
\end{center}
\caption{Simulated data showing isocontours of displacement amplitude ($\mu$m) when the attenuation and elastic modulus $M_{\infty}-M_{0}$ of the liver were varied. The viscoelastic parameters were kept constant at $M_{0}$ = 8.0 Pa and $\tau$ = 0.5 s. Blue (solid) and red (dashed) rectangles are showing the range of literature values before and after thermal ablation respectively. The blue (cross) and red (circle) markers show the average literature values for both regions.}
\label{fig:attenuation_vs_modulus}
\end{figure}

In the second part of the simulation study a Monte-Carlo type approach was undertaken to determine the effect of variations in changes in the attenuation and elastic modulus $M_{\infty}-M_{0}$ over the range of values found in the literature. The viscoelastic parameters were kept constant at $M_{0}$ = 8.0 Pa and $\tau$ = 0.5 s. The values for attenuation and elastic modulus were changed for the whole liver mimicking a therapeutic situation where multiple lesions are created next to each other. The FEM model was then used to predict the resulting ARF induced displacement with each combination of attenuation and elasticity. The isocontours at 5 $\mu$m intervals of the predicted displacement are shown in Figure \ref{fig:attenuation_vs_modulus} as a function of attenuation and elasticity. The regions before and after ablation based on the values found in the literature are shown with blue (solid) and red (dashed) rectangles respectively with average values are denoted by blue (cross) and red (circle) markers. The displacement for the average values of unablated tissue was 9 $\mu$m and this decreased by 30\% to 6 $\mu$m when employing the average values for ablated tissue.

\pagebreak

%%%%%%%%%%% DISCUSSION
\section*{Discussion}

This study addressed the utility of ARF induced displacements as a means of monitoring HIFU. Analysis of literature data indicated that changes in acoustic attenuation and elasticity with temperature will dominate changes in the displacement. The temperature dependence of the attenuation of liver reported in the literature was modelled using a polynomial fit which was able to capture the initial decrease in the attenuation and the subsequent rise \citep{damianou1997dependence, choi2011changes, jackson2014nonlinear}.

The viscoelastic properties have been reported to increase almost exponentially at temperatures above about 50 $^{\circ}$C \citep{sapin2010temperature}. DMA measurements of bovine liver were carried out in order to determine changes to the viscoelastic properties as a function of temperature. Both conventional and fractional Zener model fits to the experimental data showed changes in the viscoelastic parameters with temperature. The fractional model was found to fit the experimental data better with up to two magnitudes lower SSE when compared to conventional model, which is consistent with \citet{kiss2004viscoelastic} found the fractional Kelvin-Voigt model to predict the frequency response better when compared to the conventional model. This is because the fractional model has stronger frequency dependence which better reflects the characteristics of dynamic modulus in liver.

The motivation of this study was to investigate how these changes in acoustic and viscoelastic tissue parameters affect the ARF induced displacements during HIFU therapy. For this purpose HMI was used to monitor thermal ablation with real-time temperature data collected using an embedded thermocouple in \textit{ex vivo} liver. The temperature acquisition using this method has been shown to suffer from so-called viscous heating artefact \citep{morris2008investigation}, which can potentially lead to an error in the measured temperature value. However, this error has been shown to be significant only during the initial few seconds of the sonication \citep{dasgupta2011beam} and with the 60-second sonications employed here the effect should be less than a few degrees along the temperature axis.

During HIFU thermal ablation the displacement amplitude was observed to decrease slightly with temperature up to 45-50 $^{\circ}$C after which an increase to about 60-70 $^{\circ}$C was seen followed by a steep decrease due to thermal necrosis \citep{wall1999thermal, wright2002denaturation}. Similar observations were also reported by \citet{heikkila2010local, konofagou2012harmonic} who found the displacement amplitudes to increase before thermal ablation. The initial slight decrease in the amplitude is not always reported, but has been found to occur in some cases \citep{konofagou2012harmonic}. The increase in displacement amplitude before necrosis could potentially be due to tissue softening at low temperatures as observed by \citet{wu2001assessment} in \textit{ex vivo} bovine muscle, but this phenomenon has not been found in liver \citep{sapin2010temperature}. Furthermore, the viscoelastic measurements conducted in this and other studies \citep{kiss2009investigation} do not support this hypothesis. The slight decrease and following increase in the displacement amplitude before thermal necrosis is therefore more likely to be caused due the behaviour of attenuation with temperature which exhibits a similar trend \citep{damianou1997dependence, choi2011changes}. The changes in viscoelasticity likely do not have a significant effect until the lesion starts to grow which will only occur later in the sonication.

To better understand how these different tissue properties affect the ARF induced displacements in liver a linear viscoelastic FEM model together with nonlinear acoustic simulations were used. The simulation model took into account the temperature dependent changes in attenuation, viscoelastic parameters and lesion growth. The effect of temperature dependent sound speed and attenuation power law coefficient on displacement amplitude were found to be minimal, and thus, they were left out from the simulation model. Furthermore, the effect of thermal lensing \citep{connor2002bio} was not taken into account which could potentially shift the focal point of the ultrasound field with temperature. The magnitude of focal shift due to thermal lensing is related to the ultrasound frequency and F-number of the transducer which, for the configuration used here, should result in a few millimetre axial shift of the ultrasound focal point towards the transducer during sonication. In practice this would mean that the lesion would slightly spread in size towards the transducer. This could potentially lead to slightly lower displacement amplitudes than shown in the simulation results due to the error in the lesion size.

The simulated displacement amplitudes in liver were seen to decrease up to 40-45 $^{\circ}$C together with the attenuation where it reached its local minimum. The decrease in simulated displacement amplitude was observed to be approximately 5\% which is larger compared to experimentally observed $\sim$1\% average decrease. This is probably because the viscoelastic properties of liver were kept constant while only the value of attenuation was changed. In reality, the heating of the liver in the ultrasound focal point would cause slight changes in the viscoelasticity before the ablation, which are likely to counteract the effect of attenuation changes.

After 45 $^{\circ}$C the attenuation started to increase which also caused the displacement amplitude to increase up to 55 $^{\circ}$C. Similar increasing amplitude was also seen in the experimental data where the peak occurred approximately at 60-70 $^{\circ}$C. The difference in the peak location on temperature axis is probably due to different transducer configuration and therapeutic geometry between the experiments and simulations, which cause differences in the lesion size. However, the magnitude of the increase was similar in both the experimental and simulation data with both showing approximately 3-5\% increase. There is a large deviation in the experimental displacement values in the 60-70 $^{\circ}$C temperature region, which indicates peaking (and hence thermal ablation) does not occur at constant temperature. Therefore, it is important to record the whole time history of the displacement data to ascertain when the ablation occurs. Furthermore, there is a threshold in the lesion size (i.e., 240 CEM$_{43^{\circ}\mathrm{C}}$ volume) which starts to cause the decrease in the displacement amplitude, but its determination requires further research.

After 55 $^{\circ}$C the attenuation further increased but simultaneously a lesion started to develop causing the displacement amplitude to rapidly decrease as the increasing stiffness of the lesion outweighs the effect of increasing attenuation. The decrease in amplitude was observed to be approximately 51\% in the simulations while experimental data showed larger variation of 6-42\% decrease.

The final simulation study considered how the displacement amplitudes behave in the situation after HIFU therapy where one large lesion or multiple small lesions are next to each other. For this purpose the typical ranges in attenuation and elastic modulus in liver were used. The regions before and after thermal ablation were found to overlap slightly, although this was partly due to the relatively high elastic modulus values found in the literature for normal liver \citep{righetti1999elastographic}. Furthermore, the large deviation in the attenuation coefficient also makes the two regions to overlap. When looking at the average values the simulations predicted a 30\% decrease in displacement after ablation. However, the 5, 10 and 15 $\mu$m contours traverse both regions of non-ablated and ablated tissue, i.e., the blue and the red rectangles. This suggests that looking only at the changes in ARF induced displacement before and after thermal ablation is not enough to determine whether the tissue has undergone ablation, but rather the whole time history of the displacement during the ablation should be considered.

%%%%%%%%%%% Conclusions
\section*{Conclusions}

The effect of temperature dependent acoustic and viscoelastic tissue parameters on ARF induced displacements in liver was studied. The temperature dependent tissue properties that will most strongly affect the magnitude of ARF induced displacements were found to be attenuation and viscoelasticity. The effect of temperature dependent attenuation power law coefficient and sound speed were found to be negligible.

Experimental HMI data during HIFU ablation in \textit{ex vivo} liver showed a small but detectable decrease and increase in the displacement amplitude at low temperatures below 60-70 $^{\circ}$C followed by a more pronounced decrease due to thermal ablation. The viscoelastic simulation model, employing temperature dependent tissue parameters, exhibited similar behaviour with attenuation affecting the displacement amplitude at lower temperatures before thermal ablation. At higher temperatures, the lesion growth together with the exponentially increasing elastic modulus overpowered the changes in attenuation.

At temperatures above 60 $^{\circ}$C the experimental HMI data showed large deviation in the displacement amplitude. Both experimental data and numerical simulations indicate that monitoring displacement before and after HIFU ablation is not sufficient to determine the state of ablation. Therefore, it is concluded that monitoring ARF induced displacements continuously during thermal ablation is necessary in order to ascertain when ablation occurs.
        
%%%%%%%%%%% ACKNOWLEDGEMENTS
\section*{Acknowledgements}

V.~S. acknowledges the support of the RCUK Digital Economy Programme grant number EP/G036861/1 (Oxford Centre for Doctoral Training in Healthcare Innovation) as well as the support of Instrumentarium Science Foundation, Jenny and Antti Wihuri Foundation, Finnish Cultural Foundation, Finnish Foundation for Technology Promotion and Otto A. Malm Foundation. R.~C. acknowledges the support of EPSRC grant number EP/K02020X/1. 

%\pagebreak


%\appendix
%\section*{Appendices}

%%%%%%%%%%% REFERENCES
%% REFERENCE FORMATTING INSTRUCTIONS

%% All bibliography information should be included using a 'thebibliography' environment.  Most authors will find it easiest to create a .bbl file using the commands \bibliographystyle{} and \bibliography{} and then copy and paste the contents of the .bbl file into the .tex file below, but before the figure captions section.  Examples for using the \bibliographystyle and \bibliography commands are listed below.  

%% Do not remove the page break here.
\pagebreak

\end{document}